\documentclass{article}
\usepackage[utf8]{inputenc}
\usepackage{graphicx}
\usepackage{subcaption}
\usepackage{array}
\usepackage{adjustbox}
\usepackage{tikz}
\usepackage{bbm}
\usepackage{siunitx}
\newcolumntype{P}[1]{>{\centering\arraybackslash}m{#1}} 
\usepackage{algorithm2e}
\SetKwComment{Comment}{+-+}{+-+}
\RestyleAlgo{ruled}
\usepackage{amsmath}
\usepackage{amssymb}
\usepackage{dsfont}  
\usepackage{authblk}
\usepackage{xr-hyper}
\usepackage{hyperref}
\usepackage{upgreek}
\DeclareSIUnit{\wtpercent}{wt\%}
\usepackage{mathrsfs}
\usepackage[title]{appendix}
\usepackage{float}
\usepackage{booktabs}
\usetikzlibrary{shapes, arrows, positioning,decorations.markings}
\tikzstyle{vecArrow} = [thick, decoration={markings,mark=at position
	1 with {\arrow[semithick]{open triangle 60}}},
double distance=1.4pt, shorten >= 5.5pt,
preaction = {decorate},
postaction = {draw,line width=1.4pt, white,shorten >= 4.5pt}]
\tikzstyle{innerWhite} = [semithick, white,line width=1.4pt, shorten >= 4.5pt]
\hypersetup{hidelinks}

\usepackage[tmargin=1in,bmargin=1in,lmargin=0.75in,rmargin=0.75in]{geometry}

\usepackage{xcolor}

\newcommand\blfootnote[1]{%
	\begingroup
	\renewcommand\thefootnote{}\footnote{#1}%
	\addtocounter{footnote}{-1}%
	\endgroup
}

\usepackage[nolist,nohyperlinks]{acronym}

\usepackage{multirow}
\usepackage{makecell}

\usepackage[biblabel,nosort]{cite}

\def\submission{0}
\if\submission1
\usepackage[doublespacing]{setspace}
\usepackage{lineno}
\linenumbers
\fi

\title{Quantitative characterization of hydrophobic agglomeration at different mixing intensities using a copula-based probabilistic modeling approach}

\author{Niklas Eiermann$^{1,\ast}$, Orkun Furat$^{1,\ast}$, Jan Nicklas$^{2}$, Urs A. Peuker$^2$,  Volker~Schmidt$^1$}

\date{}

\newcommand{\euler}{e}

\newcommand{\R}{\mathbb{R}}

\newcommand{\IBc}{\mathrm{IB3}}
\newcommand{\DVa}{\mathrm{DV1}}
\newcommand{\DVb}{\mathrm{DV2}}
\newcommand{\UDVa}{\mathrm{UDV1}}
\newcommand{\UDVb}{\mathrm{UDV2}}

\newcommand{\MixingRatio}{m}
\newcommand{\abbreviationPP}{\mathrm{pp}}
\newcommand{\abbreviationAGG}{\mathrm{agg}}

\newcommand{\abbreviationEND}{\mathrm{end}}
\newcommand{\BeginOrEnd}{z}

\newcommand{\delete}{\delta}
\newcommand{\deleteSet}{\Delta}

\newcommand{\assignment}{p}

\newcommand{\iteration}{j}
\newcommand{\NumIterations}{J}
\newcommand{\kernelDensityEstimation}{k}
\newcommand{\evaluation}{l}

\newcommand{\ExpEnd}{\mathrm{end}}

\newcommand{\StarSymbol}{\ast}

\newcommand{\areaEquivalentDiameter}{d}
\newcommand{\aspectRatio}{a}
\newcommand{\pmrFunction}{\gamma}
\newcommand{\pmrMaximum}{b}
\newcommand{\pmrMinimum}{a}
\newcommand{\pmrContBerParam}{\lambda}
\newcommand{\pmrContBerFunction}{F}
\DeclareMathOperator*{\argmin}{arg\,min}
\DeclareMathOperator*{\argmax}{arg\,max}
\DeclareSIUnit{\liter}{l}

\makeatletter
\newcommand*{\addFileDependency}[1]{% argument=file name and extension
  \typeout{(#1)}
  \@addtofilelist{#1}
  \IfFileExists{#1}{}{\typeout{No file #1.}}
}
\makeatother

\begin{document}

\maketitle
\vspace{-4em}
\begin{center}
	\it
	$^1$Institute of Stochastics, Ulm University, Helmholtzstraße 18, 89069 Ulm, Germany
	\\
	$^2$Institute of Mechanical Process Engineering and Mineral Processing, Technische Universität
Bergakademie Freiberg, Agricolastraße 1, 09599 Freiberg, Germany
\end{center}
\blfootnote{$^\ast$ NE and OF contributed equally. \\ \textit{Email addresses:} niklas.eiermann@uni-ulm.de (Niklas Eiermann), orkun.furat@uni-ulm.de (Orkun Furat)
}

\begin{abstract}
	\noindent
 The agglomeration of small poorly wetted alumina particles in a stirred tank is investigated. For different experimental conditions, two bivariate probability densities for the area-equivalent diameter and aspect ratio of primary particles and agglomerates, respectively, are determined, using 2D image data from an inline camera system.  Throughout each experiment, these densities do not change since the geometries of primary particles are unaffected by the experimental conditions, while large agglomerates fragment into multiple smaller ones, which results in an equilibrium state regarding the distribution of agglomerate descriptors. Mixtures of these densities are used to model the contents of the stirred tank at each time step of the experiments. Analytical functions, whose parameters characterize the agglomeration dynamics, are fitted to the time-dependent weights of these mixtures. This enables a quantitative comparison of agglomeration processes, highlighting the impact of mixing intensity on the joint distribution of agglomerate descriptors.

\end{abstract}


\section{Introduction}
The application case comes from liquid metal filtration, since undesired inclusion particles agglomerate through turbulent collision, forming larger, multi-component particles that are easier to remove \cite{Taniguchi1996}. The increased size of agglomerated inclusion particles leads to augmented particle deposition at the inner surface of ceramic depth filters, which are utilized for the cleaning of molten metal from inclusion particles \cite{Knupfer2017}. This enables metallurgists to optimize steelmaking for cleaner and more durable high-quality products. Since metal melts are opaque and high temperatures are necessary for maintaining their molten state, it is impractical to study agglomeration processes in melts. Therefore, aqueous sodium chloride ($\mathrm{NaCl}$) solutions in combination with poorly wetted, i.e., hydrophobic, particles often serve as a model \cite{Heuzeroth2015}. In the present paper, we investigate the agglomeration of silanized alumina particles within a stirring tank filled with $0.75~\si{\mol}/\si{\liter}~\mathrm{NaCl}$ at the isoelectric point of $\si{\pH}~7.3$. The silanization of the solid surface configurates a wetting angle of $134^\circ$, which results in a poor wettability of the alumina particles by the electrolyte solution, similar to the poor wettability of non-metallic inclusion particles by molten metal \cite{Nicklas2023}. Hydrophopic agglomerates are generally larger and more stable than hydrophilic ones, as nanobubbles on the surface of hydrophobic particles form capillary bridges \cite{Knupfer2017, Nicklas2023, Gruy2005}. Besides the purification of metals, agglomeration processes hold significance in the efficient separation of suspensions into liquid and particles \cite{Van2017}. Comprehending agglomeration dynamics involves studying factors such as the turbulent flow field, particle size, wettability, viscosity, and temperature.

The aim of  the  present paper is to complement the results which have been achieved in
 \cite{Nicklas2023} for various agglomeration experiments, where  mixtures of normal distributions have been fitted to single particle descriptors, e.g. the aspect ratio, of  primary particles and agglomerates segmented from 2D image data measured in a stirred tank. More precisely, the status of agglomeration processes has been investigated at 52 time steps over the course of 47 minutes long batch agglomeration experiments. Specifically, the parameters of the underlying normal distributions of the  particle descriptors of primary particles and agglomerates, respectively, as well as the mixing  weights, i.e., the time-dependent number/volume-based fractions of agglomerates (or primary particles) of all particles in the stirred tank, have been estimated. One of the key observations made in 
 \cite{Nicklas2023} is 
  that the  parameters of the normal distributions remain rather constant throughout the 52 time steps, whereas the mixing weights significantly change over time.
  This indicates that  the descriptors of   primary particles as well as agglomerates follow fixed distributions, which do not depend on the current stage of the agglomeration process, whereas
the ratio of primary particles to agglomerates changes along the process.

Building on the results obtained in \cite{Nicklas2023}, the present paper extends the probabilistic modeling approach of hydrophobic agglomeration based on bivariate data of two-dimensional particle descriptor vectors, simultaneously considering the area-equivalent diameter and the aspect ratio of primary particles and agglomerates.
In particular,  instead of focusing on one-dimensional normal distributions, a large variety of parametric probability distributions  \cite{2020SciPy} is considered as candidates for the (marginal) distributions of single particle descriptors. Furthermore, the fact is taken into account that size and shape descriptors of particles can be correlated. For example, in some cases, small particles can be nearly spherical, whereas large particles are elongated.
Thus, in general, the joint (bivariate) probability density of area-equivalent diameter and aspect ratio is not simply the product of the corresponding marginal densities, but a more sophisticated construction based on so-called copulas \cite{Nelsen2006} is needed.
 By means of an iterative  approach, we then  fit  bivariate probability densities $f_{\abbreviationPP}^\StarSymbol$ and $f_{\abbreviationAGG}^\StarSymbol$ of area-equivalent diameter and aspect ratio for  both primary particles and agglomerates. Subsequently, these time independent densities are  used to determine the fractions $\MixingRatio_{t}^\StarSymbol $ of agglomerates for each time step $t$ of the agglomeration process.
Finally, linear combinations $(1-\MixingRatio_{t}^\StarSymbol)f_{\abbreviationPP}^\StarSymbol+\MixingRatio_{t}^\StarSymbol f_{\abbreviationAGG}^\StarSymbol$ of $f_{\abbreviationPP}^\StarSymbol$ and $f_{\abbreviationAGG}^\StarSymbol$, so-called mixtures, are  used to model the contents of the stirred tank at each time step $t$ of the experiments. Note that this mixture only depends on the fractions $\MixingRatio_{t}^\StarSymbol$, whereas the densities $f_{\abbreviationPP}^\StarSymbol$ and $f_{\abbreviationAGG}^\StarSymbol$ remain constant throughout the 52 time steps, which is based on a model assumption justified in Section~\ref{sec:desc_agglomeration_process}.

The results obtained in the present paper offer opportunities for the optimization of hydrophobic agglomeration processes. For example,  for given energy inputs of the stirrer,  
various specifications 
of process parameters
and  feeded primary particles can be considered in order 
to determine their impact on the output of the agglomeration process. In particular, in future research quantitative relationships can be derived which map process parameters and model parameters of size and shape descriptors of feeded primary particles  onto model parameters as well as the fraction of agglomerates obtained at the end of the agglomeration processes.
A key advantage of our approach is that it allows for the determination of the initial state of agglomeration, which is crucial for characterizing the entire agglomeration process. By tracking how agglomeration evolves over time, the agglomeration process can be better understood. This  insight into the temporal behavior of agglomeration processes can enable targeted adjustments to process conditions, such as stirring intensity or reactant concentrations, to achieve desired agglomerate properties.
 By means of these relationships, the following inverse problem can be investigated. Namely, for given  (desired) model parameters of size and shape descriptors of agglomerates at the end of agglomeration processes, (optimal) values of feed and process parameters can be determined which are mapped onto the predefined model parameters of agglomerates.

The rest of the present paper is organized as follows. First, in Section~\ref{sec:met}, some preliminaries are given regarding the hydrophobic agglomeration process considered in this paper, followed by a short explanation of the methods exploited for the acquisition and processing of image data. In Section~\ref{sec:copula_approach}, the  data of one of the experiments (DV1) are used as an example to explain the copula-based procedure which we apply to parametrically
model the bivariate probability distributions of particle descriptor vectors. The results obtained in this paper are presented in Section~\ref{sec.res.ult} 
and discussed in Section~\ref{sec:discussion}. Section~\ref{sec.con.clu} concludes.

\section{Some preliminaries}\label{sec:met}
\subsection{Description of the hydrophobic agglomeration process}\label{sec:desc_agglomeration_process}
This section briefly describes the particle system and experimental setup of the hydrophobic agglomeration process considered in \cite{Nicklas2023}. 
Recall that agglomeration is the adherence of particles to each other after collision \cite{Kruis1997}, where a system of particles can be subdivided into so-called primary particles and agglomerates. Therefore, in the following the term ``particle'' is used to refer to either a primary particle or an agglomerate. Note that the mentioned collision of particles requires not only a minimum level of turbulence within the liquid where the particles are situated but also a sufficient number/volume-based fraction of particles (i.e., number of particles divided by the volume of the suspension). As agglomeration reaches saturation, the number/volume-based fraction of particles decreases, causing particles to become too sparsely distributed for further collisions. This results in a minimum number/volume-based fraction of particles.

A primary particle is a single solid particle, which means that its size and shape do not change throughout the course of the agglomeration process. An agglomerate, in contrast, is a particle consisting of multiple  primary particles that adhere to each other. Since the members of a set of adhering primary particles can change during an  agglomeration experiment, the size and shape of the agglomerate associated with such a set can also change over time. However, a single agglomerate's variability in size and shape does not necessarily contradict the assumption (made in the present paper) that the agglomerates' particle descriptors follow the same probability distribution during the entire agglomeration process. 
This assumption is strengthened by the circumstance that besides the growth of agglomerates,   large agglomerates can be redispersed by turbulence into multiple smaller ones, which results in an equilibrium state regarding the distribution of agglomerate descriptors.

\subsubsection{Experimental setup}\label{sec:experimental_setup}

\begin{figure}[h]
    \centering
    \includegraphics[width=0.35\textwidth]{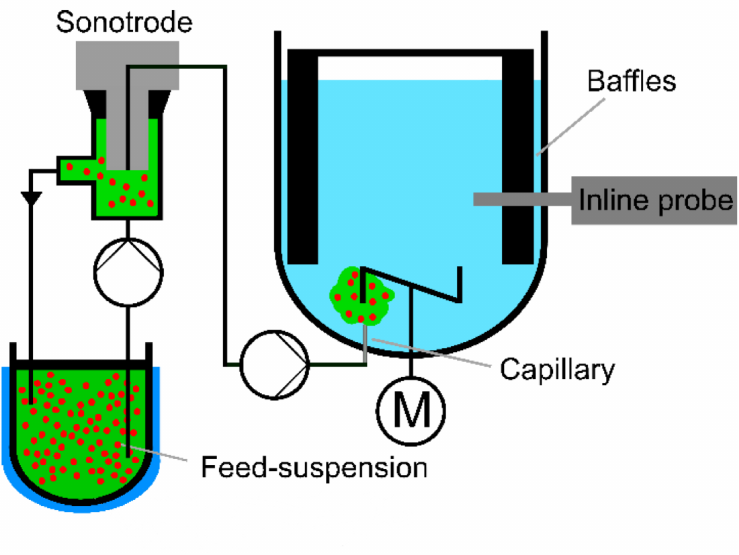}
    \caption{Experimental setup  of the hydrophobic agglomeration process; taken from \cite{Nicklas2023}.
    }
    \label{fig:experimental_setup}
\end{figure}

In \cite{Nicklas2023} the agglomeration of hydrophobic fines in a stirred tank was investigated by inline probe imaging as illustrated in  Figure~\ref{fig:experimental_setup}.

\begin{figure}[h!]
  \centering
  
  \begin{minipage}[c]{0.35\textwidth}
    \centering
    \begin{adjustbox}{width=0.67\textwidth,center,raise=0pt}
      \includegraphics{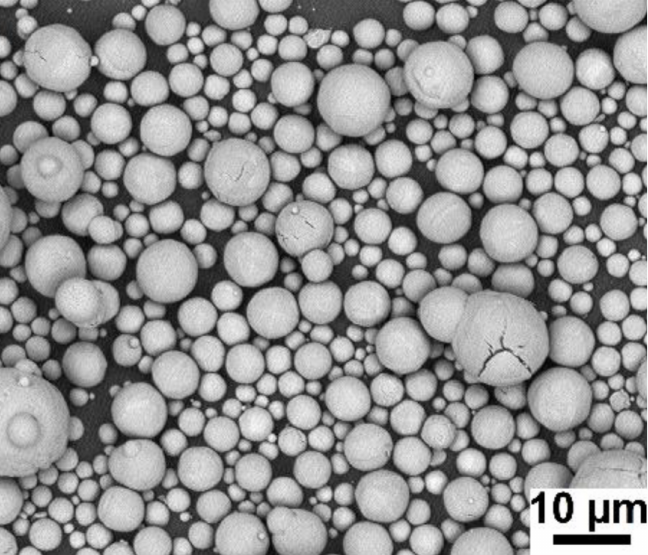}
    \end{adjustbox}
  \end{minipage}\hspace{0.5cm}
   \begin{minipage}[c]{0.35\textwidth}
    \centering
    \begin{adjustbox}{width=\textwidth,center}
      \includegraphics{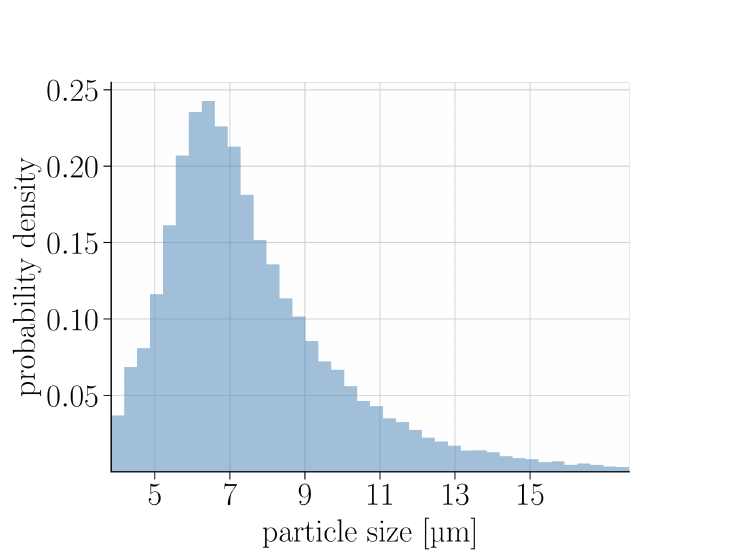}
    \end{adjustbox}
  \end{minipage}
  \caption{SEM image of silanized alumina particles (left), and histogram of the
  number-based particle size distribution from inline probe data for primary particles dispersed in pure ethanol (right);
  taken from  \cite{Nicklas2023}.}
  \label{fig:particles_in_ethanol}
\end{figure}

The experimental setup in Figure~\ref{fig:experimental_setup} shows a ``feed-suspension'', which consists of silanized alumina particles and ethanol. The silanization ensures the hydrophobicity of the particles, whereas the ethanol functions as an organic solvent for dispersing the particles into almost exclusively primary particles.  In Figure~\ref{fig:particles_in_ethanol}   a scanning electron microscopy (SEM) image is displayed, which highlights the almost perfectly spherical shapes of the primary particles. as well as 
 a histogram of the number-based particle size distribution.

The tank shown in Figure~\ref{fig:experimental_setup} is equipped with four baffles and is, prior to the addition of the ``feed-suspension'', solely filled with $\num{3.5}~\unit{\liter}$ of $0.75~\mathrm{mol/l}$ $\mathrm{NaCl}$ solution with a $\si{\pH}$-value adjusted to the isoelectric point of $\si{\pH}~7.3$ for silanized alumina particles \cite{Nicklas2023}. In the experiments analyzed in the present paper, two different stirrer designs, both with $5$ cm diameter, are considered.

The inclined blade (IB) stirrer uses four blades that are inclined by 45 degrees. This stirrer leads to the generation of an axial flow of the liquid from the top towards the bottom and subsequently recirculates the liquid along the walls of the tank. In contrast, the use of the Rushton-turbine, which is equipped with six upright blades, leads to the generation of axial flows from both, below and above the stirrer, and forwards them in radial directions. For reaching a constant stirrer speed of $450~\mathrm{rpm}$ that is used across all experiments considered in this paper, the Rushton turbine requires a three times higher power input than the IB stirrer. As a consequence, the flow induced by the Rushton-turbine results in a three times higher (mass averaged) energy dissipation rate $\varepsilon$ in comparison to the flow induced by the IB stirrer \cite{Wang2014}, see Table~\ref{tab:experiments}. Note that $\varepsilon$ is determined by dividing the power input of the stirrer running at $450~\mathrm{rpm}$ by the mass of the liquid in the tank, which is constant across all considered experiments in Table~\ref{tab:experiments} that describes the  three different configurations of the  experiments discussed in this paper.

\begin{table}[h]
    \centering
        \begin{tabular}{P{2.1cm}|P{2.9cm}|P{1cm}|P{1.9cm}} 
            experiment & stirrer & $\varepsilon$ in $\mathrm{W/kg}$ & ultrasound\\ \specialrule{2pt}{0em}{0em}
            $\DVa$/$\DVb$ & Rushton-turbine & $0.3$ & yes\\ \hline
            $\UDVa$/$\UDVb$ & Rushton-turbine & $0.3$ & no\\ \hline
            $\IBc$ & inclined blade (IB) & $0.1$ & yes
        \end{tabular}
    \caption{Configuration of   experiments, where $\DVb$ and $\UDVb$ are repetitions of $\DVa$ and $\UDVa$, respectively.}
    \label{tab:experiments}
\end{table}

Specifically, the considered experiments are denoted by $\DVa$, $\DVb$, $\UDVa$, $\UDVb$ and $\IBc$. Here, DV means ``dispersed version'', and UDV stands for  ``undispersed version'', indicating whether ultrasound was applied, i.e., whether the sonotrode was running in the dispersion loop, to further disperse the particles contained in the ``feed-suspension'' that had already been dispersed to some extent by ethanol.
By $\DVb$ and $\UDVb$, repetitions of $\DVa$ and $\UDVa$ are denoted. 
This means that the ratio of agglomerates to primary particles at the beginning of the experiments without ultrasonic dispersion of preexisting agglomerates, namely $\UDVa$ and $\UDVb$, should be higher than for the other experiments listed in Table~\ref{tab:experiments}.

\subsubsection{Breakage behavior of agglomerates}

The energy dissipation rate $\varepsilon$ mentioned in Section~\ref{sec:experimental_setup} quantifies the amount of energy that is transferred from the stirrer to the liquid inside the stirred tank and thus serves as a measure for the energy available for the generation of turbulence. Therefore, the energy dissipation rate is positively correlated with the number of occurring particle collisions. Furthermore, the energy dissipation rate $\varepsilon$ is relevant because of its influence on the Kolmogorov microscale $\eta$ of the turbulence, which is given by
\begin{equation}
    \eta = \bigg(\frac{\nu^3}{\varepsilon}\bigg)^{1/4},
\end{equation}
where $\nu$ is the kinematic viscosity. Particles that are larger than $\eta$ are exposed to greater shear forces that can cause agglomerates to redisperse into multiple smaller particles \cite{Knupfer2017}. This means that for a larger $\varepsilon$ smaller agglomerates can break due to shear forces. Therefore, $\eta$ represents the approximate maximum size of agglomerates under turbulence.

\subsection{Acquisition and processing of image data}\label{sec:acq_data}

In this section we briefly recall the methods used in \cite{Nicklas2023} for the acquisition and processing of image data. 

\subsubsection{Inline camera system} 

\begin{figure}[h]
  \centering
   \begin{minipage}[c]{0.35\textwidth}
    \centering
    \begin{adjustbox}{width=0.8\textwidth,center}
      \includegraphics{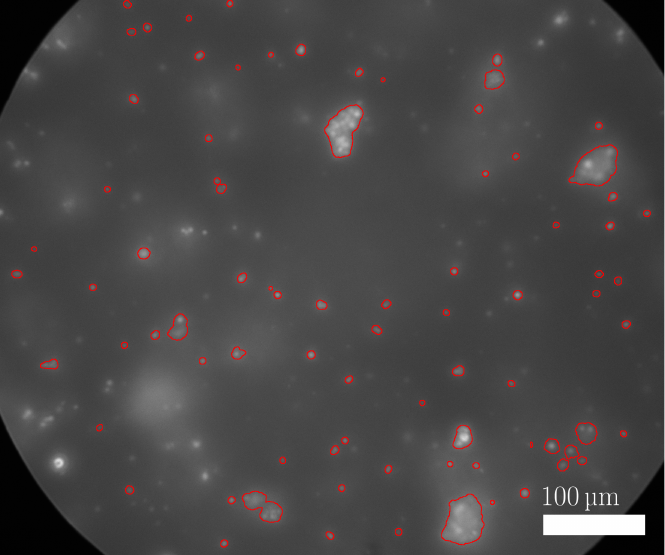}
    \end{adjustbox}
  \end{minipage}\hspace{1cm}
  \begin{minipage}[c]{0.38\textwidth}
    \centering
    \begin{adjustbox}{width=0.99\textwidth,center,raise=0pt}
      \includegraphics{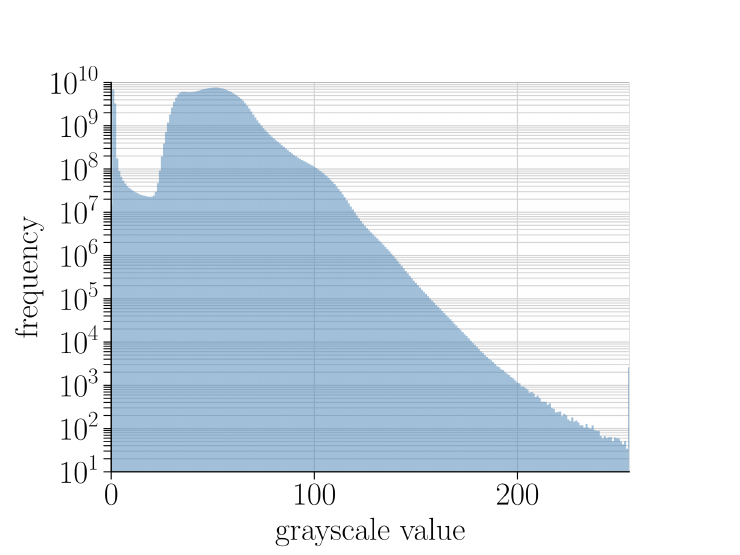}
    \end{adjustbox}
  \end{minipage}
  \caption{Contrast enhanced image (left) captured with SOPAT PL with segmented particles outlined in red, histogram (right) of grayscale values of the pixels in all images captured for the  experiments  listed in Table \ref{tab:experiments}. }
  \label{fig:segmented_image}
\end{figure}

Primary particles and agglomerates in the stirred tank were captured with the inline camera probe SOPAT PL (SOPAT GmbH, Germany). The probe's immersible shaft of $\SI{12}{\milli\meter}$ diameter reaches $\SI{3.5}{\centi\meter}$ into the stirred tank. A built-in strobe allows for a good illumination of even the smallest particles. At the end of the probe a rhodium reflector is positioned at a distance of $\SI{4}{\milli\meter}$ to the camera lens. The large distance to the lens ensures that the particles in the stirred tank are sampled representatively. The camera system captures 8-bit grayscale images with a resolution of $2464\times 2056$ square-shaped pixels, where each pixel has a side length of $\SI{0.2646}{\micro\meter}$. Figure~\ref{fig:segmented_image} (left) shows a typical captured image, where segmented particles are outlined in red. Note that pixels that belong to particles tend to have larger grayscale values than pixels that belong to the background. In addition, Figure~\ref{fig:segmented_image} (right) shows the distribution of grayscale values of the pixels in a total of 
52,000 images captured for the  experiments  listed in Table \ref{tab:experiments}.

The following 47 minutes long procedure is performed for each  experiment  considered in this paper. First, the suspension is added to the tank with the stirrer already running. After a mixing period of 30 seconds, the image acquisition starts at time step $t=1$. Then, within ten seconds, the camera system captures 200 images. After those ten seconds and an additional waiting period of 45 seconds, i.e., 55 seconds after $t=1$, time step $t=2$ is reached. For $t=2$, the same imaging procedure as for $t=1$ is performed. Overall, this procedure is repeated for 52 time steps, where  200 images are acquired  for each time step $t\in T=\{1,2,\dots,52\}$.

\subsubsection{Segmentation algorithm}\label{sec:segmentation}

This section summarizes the image preprocessing and segmentation steps performed in \cite{Nicklas2023}, where as a first step, a few undesirable effects  occurring in the image data had to be removed. In some images, particles are observed that deposit directly on the lens of the probe, which appear relatively large. Moreover, these particle depositions may remain on the lens for multiple frames. Since considering these particles would lead to a non-representative particle distribution, it is desirable to skip them in the segmentation algorithm. Furthermore, there are particles that appear to be out of the focal plane. Similar as the depositions on the lens they show rather blurry edges, compare Figure~\ref{fig:segmented_image} (left).

The initial step of the segmentation algorithm involves applying a bilateral Gaussian filter \cite{Tomasi1998} to achieve image smoothing. For each pixel, this filter computes a convex combination of neighboring pixel values within a predefined window. The weights in these convex combinations depend on the distance between the pixels, taking into account both their spatial positions and grayscale value differences. The inclusion of grayscale value differences in the filter aims to preserve edges.

Even though particles appear to be brighter than the background, binarizing images with a single threshold is not suitable for reliably segmenting the particles. This is due to significantly different grayscale values of pixels associated with particles across different areas of the same image which is the result of variations in particle concentration, out-of-focus particles, and particles being attached to the lens. To address this, a range filter is applied to the image. In doing so, each pixel of the image is assigned the span of values of all neighboring pixels in a $17\times 17$ window around the pixel. Thus, this filter enhances particle edges while disregarding out-of-focus particles. This aims to distinguish particles in focus from out-of-focus particles and background.
The resulting range-filtered image is then thresholded by setting pixels below a threshold of 9 to zero, where the pixels of unprocessed 8-bit  images have grayscale values in $\{0,1,\dots,255\}$, see Figure~\ref{fig:segmented_image}. After  thresholding, a morphological closing operation \cite{Gonzales2010} aims to remove small connected components of non-zero pixels with 50 pixels or fewer, where the structuring element  has a circular shape with a diameter of 119 pixels. The remaining connected components of non-zero pixels represent regions of interest (ROIs) for further processing.

To perform binarization, each identified ROI is processed separately. For a reliable segmentation of particles in a ROI, a combination of two binarization techniques was used: global binarization with an image specific constant threshold and a local binarization method, which computes local thresholds in the ROI based on local mean grayscale values. The results from both methods are combined by setting a pixel value to one if it exceeds at least the lower threshold of the two techniques. On the one hand, this combination is necessary due to the possibly large range in the unobservable $z$-direction in which larger agglomerates extend resulting in a wide range of grayscale values. For such particles, the local binarization method is inapt and a global binarization method is required. On the other hand, the global binarization method's identification of edges of blurry objects can be unprecise. Thus, neither method is suitable to be utilized independently. 

\begin{figure}[h]
  \centering
  \begin{subfigure}{0.35\textwidth}
    \includegraphics[width=\textwidth]{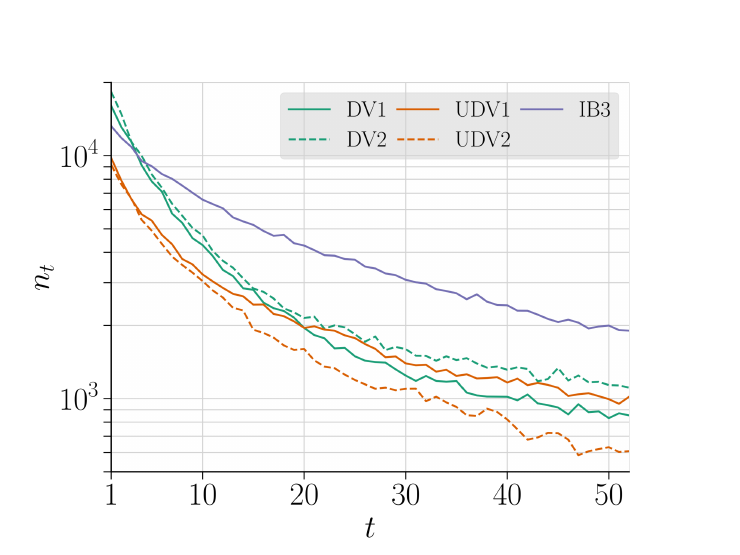}
    \caption{}
  \end{subfigure}
  \hspace{1cm}
  \begin{subfigure}{0.35\textwidth}
    \includegraphics[width=\textwidth]{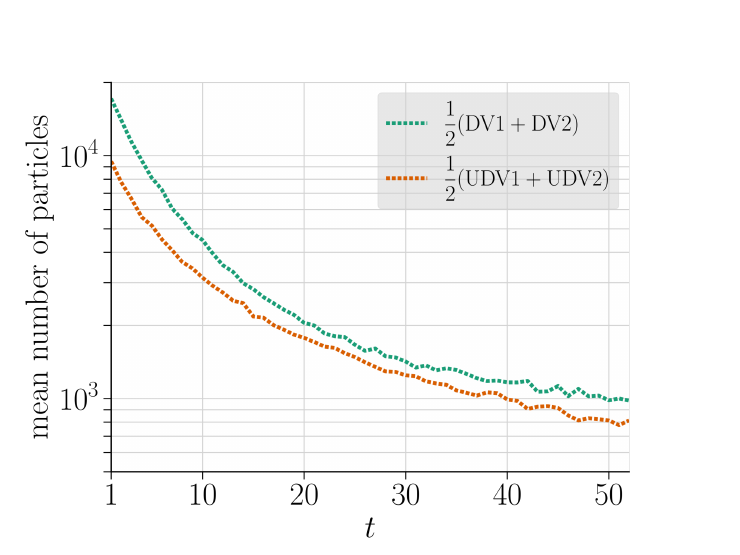}
    \caption{}
  \end{subfigure}
    \caption{ (a) Number $n_t$ of  particles segmented at  time step $t\in T$ for the experiments DV1, DV2, UDV1, UDV2 and IB3.  (b)  Mean number of particles of experiments $\DVa$ and $\DVb$  (green) as well as $\UDVa$ and $\UDVb$  (orange) at  time step $t\in T$.}
    \label{fig:number_of_particles}
\end{figure}

For each of the five experiments listed in Table~\ref{tab:experiments},
Figure~\ref{fig:number_of_particles} shows the number of particles $n_{t}$ extracted from the grayscale images from the corresponding time step $t\in T$. Note that the decreasing number of particles in all experiments is not only caused by agglomeration where multiple primary particles are combined to a single multi-component particle. There is a small quantity of particles, mostly the larger ones, that adhere to the baffles of the tank or immobilize at the gas-liquid interface at the top of the suspension in the stirred tank, see Figure \ref{fig:experimental_setup}, where they are not captured by the inline probe.

\section{Stochastic modeling of particle descriptor vectors}\label{sec:copula_approach}

This section uses the data  of experiment $\DVa$ as an example to explain the procedure which we apply  to parametrically model the probability distributions of  particle descriptor vectors for each time step $t\in T$.  First, in Section~\ref{sec:geometric_descriptors}, we introduce the geometrical particle descriptors considered in this paper. In Section~\ref{sec.cop.mod}, we provide some  fundamentals of the copula-based approach which we exploit to model the joint distribution of descriptor vectors, where we begin with  modeling the marginal distributions of individual components. More precisely, we fit univariate parametric probability densities to the area-equivalent diameter and aspect ratio of the  particles measured in experiment $\DVa$ at time step $t=1$, see Section~\ref{sec:mod_single}.
Then, in Section~\ref{sec.two.dim} the joint (bivariate) probability density of these two particle descriptors is modeled by means of so-called Archimedean copulas. The use of this class of copulas has the advantage that associated copulas typically have relatively few parameters to describe bivariate densities while offering a broad range of different copula types to flexibly capture various dependency structures. 
A small number of parameters has the advantage of preventing dependency structures from being overfitted, especially when only a limited amount of data is available.
In comparison, non-parametric methods for modeling bivariate densities, such as kernel density estimation, can be deployed. However, since they are more flexible than copula-based approaches, they also require more data for fitting purposes---in particular to avoid overfitting \cite{Furat2019}. 

The copula-based fitting of an (initial) bivariate density  for descriptor vectors of agglomerates  is explained in Section~\ref{sec:FittingModelParams}, whereas Section~\ref{sec:iterative} deals with
iterative adjustments of the fitted bivariate probability densities.
Finally, 
the model validation considered in Section~\ref{sec:model_validation} provides tools for evaluating the quality of the developed model.

\subsection{Geometrical particle descriptors}\label{sec:geometric_descriptors}
In the following, we will investigate the size and shape of particles extracted from grayscale images as described in Section~\ref{sec:acq_data}. For that purpose, we consider the area-equivalent diameter $d(P)$ and aspect ratio $a(P)$ of the planar projection   $P\subset\mathbb{R}^2$ of  three-dimensional particles.\footnote{From now on, for brevity,  $P$ is called a particle, instead of saying ``planar projection'' of a particle.}

Recall that the area-equivalent diameter $d(P)$ of  $P$ is given by
\begin{equation}
    d(P)=2\sqrt{\frac{\mathrm{area}(P)}{\pi}},
\end{equation}
where $\mathrm{area}(P)>0$ denotes the area of $P$ (in \si{\micro\meter\squared}). Furthermore, the aspect ratio $a(P)\in [0,1]$ of $P$ will be considered, which may be interpreted as a measure of roundness. In particular, an aspect ratio close to one indicates that a particle has similar width in all directions, whereas an aspect ratio close to zero corresponds to a particle with an elongated shape in one direction compared to the others. To formally define the aspect ratio $a(P)$, we use the Feret diameters of $P$, see \cite{Al-Thyabat2006}. For each pair of parallel tangents, which are touching opposite sides of a particle's outline, the Feret diameter is given by the perpendicular distance between these tangents. 
The aspect ratio is then given by
\begin{equation}
    a(P)=\frac{\mathrm{feret}_{\min}(P)}{\mathrm{feret}_{\max}(P)},
\end{equation}
where $\mathrm{feret}_{\min}(P)$ and $\mathrm{feret}_{\max}(P)$ denote the the minimum and maximum Feret diameters of $P$, respectively.

\subsection{Copula-based modeling approach}\label{sec.cop.mod}

\begin{figure}[h]
  \centering
  \begin{subfigure}[b]{0.45\textwidth}
    \includegraphics[width=\textwidth]{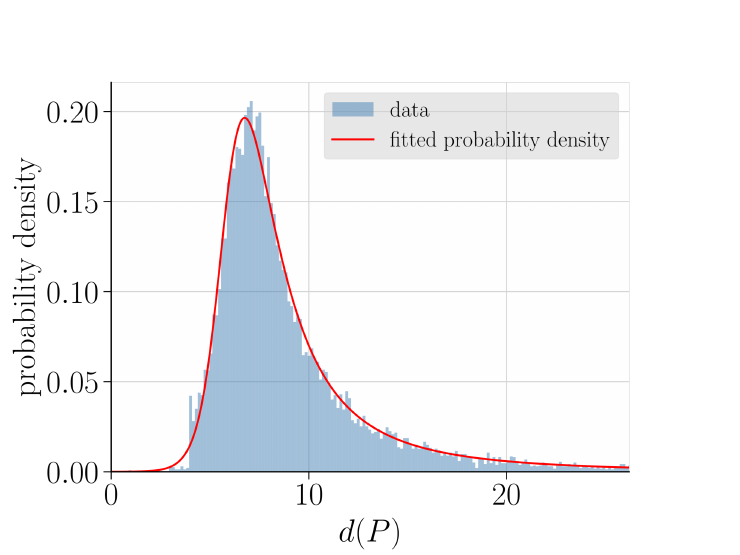}
    \caption{}
  \end{subfigure}
  \hspace{1cm}
  \begin{subfigure}[b]{0.45\textwidth}
    \includegraphics[width=\textwidth]{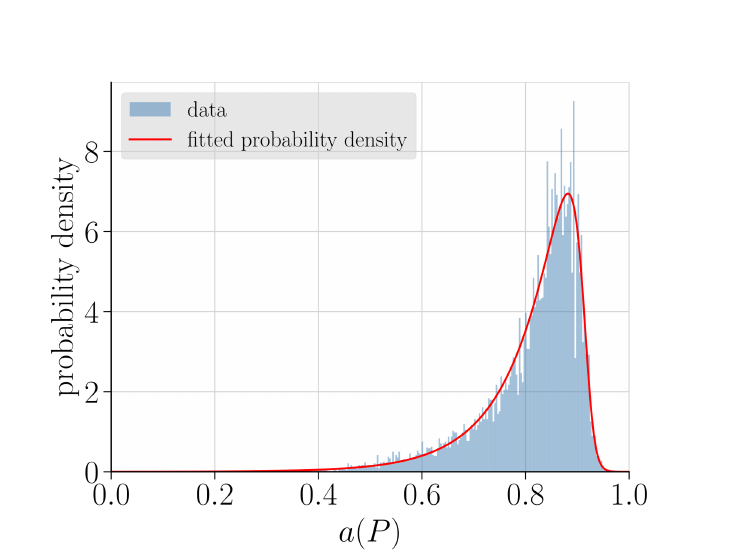}
    \caption{}
  \end{subfigure}
  \caption{Parametric fits (red curves) for the univariate probability densities of  area-equivalent diameter (a) and  aspect ratio (b) at time step $t=1$ for experiment $\DVa$. }
  \label{fig:single_fit}
\end{figure}

As already mentioned above, the first measurement at time step $t=1$ shows a dominating share of primary particles. In fact, histograms of the particle descriptors $d(P)$ and  $a(P)$  obtained from this measurement indicate unimodal probability densities, see Figure~\ref{fig:single_fit}, whereas the presence of a larger fraction of agglomerates would lead to bi- or even multi-modality.
Therefore, as an initial approach for $t=1$, it seems to be reasonable to model the distributions of these descriptors by unimodal densities. 
However, beginning from Section~\ref{sec:FittingModelParams} we will consider linear combinations of two unimodal densities for the descriptors of mixtures of primary particles and agglomerates.

\subsubsection{Univariate distribution of single particle descriptors}\label{sec:mod_single}

To get an initial model for the univariate distributions of  $d(P)$ and  $a(P)$ at time step $t=1$, we consider parametric families of probability distributions from an extensive list of candidates \cite{2020SciPy} that includes, for example, the normal distribution. Furthermore, in order to determine suitable parameter values of such distributions we use maximum likelihood estimation \cite{Held2014}. As a result,   for each parametric family  we get a distribution that best fits, with a corresponding likelihood. Then, we choose the parametric family with the largest likelihood as model for the distribution of the descriptor under consideration.

Among the candidate distributions provided in \cite{2020SciPy}, the so-called generalized hyperbolic distribution  scored the largest likelihood for both descriptors,  $d(P)$ and  $a(P)$,  
where the density $f_{\alpha, \beta, \delta, \lambda, \mu}:\R\to\R$ of the generalized hyperbolic distribution  is given by
\begin{equation}\label{eq.gen.par}
    f_{\alpha, \beta, \delta, \lambda, \mu}(x)=\frac{\euler^{\beta(x-\mu)}}{\alpha^{\lambda-\frac{1}{2}}\delta^\lambda\sqrt{2\pi}}(\delta^2+(x-\mu)^2)^{\lambda-\frac{1}{2}}(\alpha^2-\beta^2)^{\lambda/2}\frac{\mathrm{B}_{\lambda-\frac{1}{2}}(\alpha\sqrt{\delta^2+(x-\mu)^2})}{ \mathrm{B}_\lambda(\delta\sqrt{\alpha^2-\beta^2})},
\end{equation}
for each $x\in \R$, see \cite{barndorff1977}. Here $(\beta, \lambda, \mu)\in\mathbb{R}^3$, $(\alpha, \delta)\in (0,\infty)^2$ with $|\beta|<\alpha$ are model parameters. 
After fitting the parameters $\alpha, \beta, \delta, \lambda, \mu$ to data and inserting them into the right-hand side of Eq.~(\ref{eq.gen.par}), the values of $f_{\alpha, \beta, \delta, \lambda, \mu}(x)$ can be computed, see
the red curves in Figure~\ref{fig:single_fit}, where  two evaluations of the modified Bessel function $\mathrm{B}_q$ of the third kind with index $q$ \cite{Abramowitz1968} are required. 

For the first time step $t=1$ of experiment DV1, considered in this section as an example, 
we obtained the following  values for the model parameters $\alpha, \beta, \delta, \lambda, \mu$. Namely, $\alpha=0.7192, \beta=0.6848, \delta=2.2314, \lambda=-1.0307, \mu=5.7799$
for the area-equivalent diameter $d(P)$, and $\alpha=81.5848, \beta=-72.0746, \delta=0.0301, \lambda=0.5085, \mu=0.9208$
for the aspect ratio $a(P)$.

\subsubsection{Bivariate distribution  of two-dimensional particle descriptor vectors}\label{sec.two.dim}

If the particle descriptors $d(P)$ and  $a(P)$ introduced in Section~\ref{sec:geometric_descriptors} could be considered to be  independent random variables, the bivariate probability density 
$f:\R^2\to[0,\infty)$ of the two-dimensional descriptor vector $(d(P),a(P))$ would be given by 
\begin{equation}\label{for.pro.den}
f (x_1,x_2)= f_{\areaEquivalentDiameter}(x_1) f_{\aspectRatio}(x_2)\qquad\mbox{ for any $x_1,x_2\in\R$},
\end{equation}
where $f_\areaEquivalentDiameter$ and $f_\aspectRatio$ denote the univariate probability densities of $d(P)$ and $a(P)$, respectively. However, this approach is not suitable for correlated descriptors, as it is the case for the descriptors  $d(P)$ and  $a(P)$  of experiment $\DVa$ at time step $t=1$. Here, we obtained a Pearson correlation coefficient of  $-0.4$ for  $d(P)$ and  $a(P)$, which is clearly distinct from zero.
 
One way of modeling the joint distribution of correlated particle descriptors is through multivariate normal distributions~\cite{Anderson2003}, with corresponding correlation matrices that capture the particle descriptors' interdependencies, like in  \cite{Nicklas2023}. 
However, the marginals of multivariate normal distributions are (univariate) normal distributions, which are symmetric and therefore inadequate to model our datasets, whose histograms are skewed,  see Figure~\ref{fig:single_fit}. 
As shown in Section~\ref{sec:mod_single},  other families of parametric probability distributions, like  generalized hyperbolic distributions, are more suitable for modeling the
distributions of $d(P)$ and  $a(P)$. To come up with a parametric model for the joint distribution of   the two-dimensional descriptor vector $(d(P),a(P))$, we consider so-called Archimedean copulas \cite{Nelsen2006}, which  have been successfully exploited in previous studies of data for particle descriptor vectors,  see e.g. \cite{Furat2019}. Specifically, copulas allow us to incorporate both flexibility in choosing the families of marginal distributions and accurate modeling of correlations between particle descriptors into our model.

\paragraph{Sklar's representation formula.}\label{sec:basic_concepts}
 Note that the product formula given in Eq.~(\ref{for.pro.den}) is a special case of the following representation formula for bivariate probability densities, which is a differential version of  Sklar's representation formula for cumuluative distribution functions,  see e.g. \cite{Nelsen2006}.
 Namely, for  the  bivariate density 
$f:\R^2\to[0,\infty)$ of the two-dimensional descriptor vector $(d(P),a(P))$ it holds that
\begin{equation}\label{for.cop.den}
f(x_1,x_2)= c(F_d(x_1),F_a(x_2)) f_{d}(x_1) f_{a}(x_2)\qquad\mbox{ for any $x_1,x_2\in\R$},
\end{equation}
where $F_d:\R\to[0,1]$ and $F_a:\R\to[0,1]$ are the cumulative distribution functions of $d(P)$ and $a(P)$, respectively, and $c:[0,1]^2\to[0,\infty]$ is a so-called copula density, which is the bivariate probability density of a two-dimensional random vector $(U_1,U_2)$ such that its components $U_1$ and $U_2$ are uniformly distributed random variables with values in the unit interval $[0,1]$. In particular, in the independent case considered in  Eq.~(\ref{for.pro.den}) it holds  that $c(u_1,u_2)=1$ for any $u_1,u_2\in[0,1]$.

\paragraph{Archimedean copulas.}\label{sec:Archimedean_copulas}
To parametrically model
the copula density $c:[0,1]^2\to[0,\infty]$ appearing in Eq.~(\ref{for.cop.den}), we consider the cumulative distribution function $C:[0,1]^2\to[0,1]$ of $(U_1,U_2)$ given by
\begin{equation}\label{cum.dis.fun}
C(u_1,u_2)=\int_0^{u_1}\int_0^{u_2} c(v_1,v_2) {\areaEquivalentDiameter} v_2{\areaEquivalentDiameter}v_1\qquad\mbox{for any $u_1,u_2\in[0,1]$.}    
\end{equation}
Note that $C$ is called the copula corresponding to the copula density $c$. In this paper we consider a special class of copulas, so-called Archimedean copulas, whose definition is based on  Archimedean generators $\varphi \colon [0,1]\to [0,\infty]$, which are continuous, strictly decreasing and convex functions with $\varphi(0)=\infty$ and $\varphi(1)=0$, see e.g. \cite{Nelsen2006}. The copula $C\colon [0,1]^2\to [0,1]$ is then given by
\begin{equation}\label{eq:ArchimedeanDefinition}
C(u_1,u_2)=\varphi^{-1}(\varphi(u_1)+\varphi(u_2)) \qquad \mbox{ for any $u_1,u_2\in [0,1]$.}
\end{equation}
Thus, to parametrically model  
the copula density $c$ appearing in Eq.~(\ref{for.cop.den}),  we consider various  parametric families $\{\varphi_\theta: \theta\in\Theta\}$ of Archimedean generators, where $\Theta\subset\R$ is some set of admissible parameters, see Table~\ref{tab:copula_families}.
Then, each  family of generators  $\{\varphi_\theta: \theta\in\Theta\}$ induces a parametric family of copula densities 
$\{c_\theta: \theta\in\Theta\}$, where
\begin{equation}\label{par.cop.den}
c_{\theta}(u_1,u_2)=\frac{\partial^2}{\partial u_1\, \partial u_2} \varphi_\theta^{-1}(\varphi_\theta(u_1)+\varphi_\theta(u_2))\qquad\mbox{ for any $u_1,u_2\in [0,1]$ and $\theta\in\Theta$.}
\end{equation}

\begin{table}[H]
	\centering
        \begin{tabular}{P{1.5cm}!{\vrule width 2pt}P{2.5cm}|P{2.5cm}|P{2.5cm}|P{2.5cm}|P{2.5cm}} 
         copula    & Ali-Mikhail-Haq & Clayton & Frank & Gumbel & Joe\\ \hline
            $\Theta$ & $[-1,1)$ & $(0,\infty)$ & $\R\setminus\{0\}$ & $[1,\infty)$ & $[1,\infty)$ \\ \hline
            $\varphi_\theta (u)$ & $\ln\frac{1-\theta(1-u)}{u}$ & $\frac{1}{\theta}(u^{-\theta}-1)$ & $-\ln\frac{\euler^{-\theta u}-1}{\euler^{-\theta}-1}$ & $(-\ln u)^\theta$ & $-\ln(1-(1-u)^\theta)$
        \end{tabular}
    \caption{Parametric families $\{\varphi_\theta: \theta\in\Theta\}$ of Archimedean generators.}
\label{tab:copula_families}
\end{table}

\paragraph{Rotated Archimedean copulas.}
For each of the  
 parametric families of copula densities 
$\{c_\theta: \theta\in\Theta\}$ introduced in Eq.~(\ref{par.cop.den}), we additionally consider three rotated versions, which are obtained by rotations of $c_\theta$ around the midpoint $(0.5,0.5)$ of the square $[0,1]^2$ by multiples of 90 degrees. In this way, we obtain a two-parametric set of copula densities $\{c_{(\theta,r)}: \theta\in\Theta,r\in R\}$, where $R=\{0,90,180,270\}$.
Inserting these copula densities into Eq.~(\ref{for.cop.den}), we obtain the parametric bivariate density 
$f_{(\theta,r)}:\R^2\to[0,\infty)$ of the two-dimensional descriptor vector $(d(P),a(P))$ for each $(\theta,r)\in\Theta\times R$, where
\begin{equation}\label{eff.the.err}
f_{(\theta,r)} (x_1,x_2)= c_{(\theta,r)} (F_d(x_1),F_a(x_2)) f_{d}(x_1) f_{a}(x_2)\qquad\mbox{ for any $x_1,x_2\in\R$.}
\end{equation}

\begin{figure}[h!]
    \centering
    \includegraphics[width=0.7\textwidth]{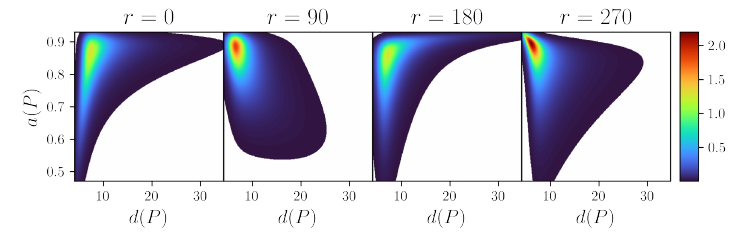}
    \caption{ Bivariate density 
$f_{(\theta,r)}$  of  $(d(P),a(P))$
 for differently rotated  Clayton copula densities $c_{(\theta,r)}$.
}\label{fig:copula_rotations}
\end{figure}

Figure~\ref{fig:copula_rotations} illustrates some effects, which can occur for differently rotated  Clayton copula densities $c_{(\theta,r)}$, where we put $\theta=1$, and for the model parameters $\alpha, \beta, \delta, \lambda, \mu$ of the  (generalized hyperbolic) marginal distributions we used the values stated in Section~\ref{sec:mod_single}.

\paragraph{Selection of the best fitting copula.}
In Section~\ref{sec:mod_single}, we selected the best fitting univariate marginal distributions for $d(P)$ and $a(P)$, respectively, by evaluating multiple parametric distribution families using maximum likelihood estimation. Here, we follow a similar approach to select the best fitting combination of an Archimedean copula type (with its corresponding parameter $\theta\in\Theta$) given in Table~\ref{tab:copula_families} along with a rotation angle $r\in R$, again based on maximum likelihood estimation. 

More precisely, we consider the bivariate
probability density $f_{(\theta,r)}$
given in Eq.~(\ref{eff.the.err}),
where the densities $f_d, f_a$ (and their corresponding distribution functions $F_d,F_a$) on the right-hand side have already been fitted, see Section~\ref{sec:mod_single}. Then, the parameter $\theta$ can be determined by means of maximum likelihood estimation, i.e., by 
\begin{equation}\label{eq:argmax_likelihood}
\widehat{\theta}=\argmax_{\theta\in\Theta} \mathcal{L}(\theta ; D_1),
\end{equation}
where 
the likelihood function  $\mathcal{L}(\theta ; D_1)$ is given by
\begin{equation}
\mathcal{L}(\theta ; D_1)= \prod_{x\in D_1} f_{(\theta,r)}(x) \qquad \text{ for each } \theta \in \Theta,\label{eq:Like}
\end{equation}
and
$D_1\subset[0,\infty)\times[0,1]$ denotes the set of all values $x=(x_1,x_2)$ available for  the descriptor vector $(d(P),a(P))$ at time step $t=1$ of the experiment under consideration (i.e., DV1 in this section).
 Then, we choose the parametric copula type  and the rotation angle, which lead to the largest likelihood
 $\mathcal{L}(\widehat\theta ; D_1)$,
 as modeling component for the bivariate distribution of  $(d(P),a(P))$.

\begin{figure}[h!]
    \centering
    \includegraphics[width=0.7\textwidth]{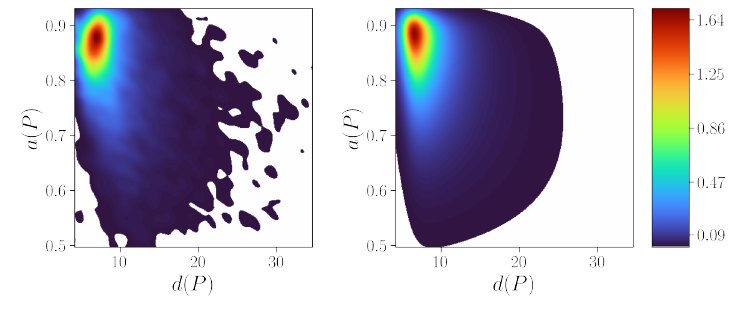}
    \caption{Bivariate probability density of $(d(P),a(P))$ obtained for time step $t=1$ of experiment $\DVa$ via
kernel density estimation (left) and the parametric copula-based fitting procedure described above (right).}
    \label{fig:t1_fit}
\end{figure}

 For the experiment $\DVa$,  the Clayton copula with $\theta=0.602$ and $r=90$ provided the best fit among the Archimedean copula types given in Table~\ref{tab:copula_families}, see also Figure~\ref{fig:t1_fit}.

\subsection{Bivariate distribution of particle descriptor vectors for agglomerates}\label{sec:FittingModelParams}

In Section~\ref{sec.cop.mod} we fitted a bivariate probability density  of the two-dimensional particle descriptor vector $(d(P),a(P))$ to data of the first measurement at time step $t=1$.
However,  the  goal of this paper is to develop a parametric model for the  joint density  of  $(d(P),a(P))$ for each time step $t\in T$, where from now on we will use the notation $f_t$ for this density, instead of  $f$ $ (=f_1)$ used so far.
More precisely, for each $t\in T$ we want to represent the (time-dependent) probability density $f_t:\R^2\to[0,\infty)$ of $(d(P),a(P))$ in the following form:   
\begin{equation}\label{eq:f_mix}
    f_{t} = (1-\MixingRatio_{t})f_{\abbreviationPP}+\MixingRatio_{t} f_{\abbreviationAGG},
\end{equation}
where $\MixingRatio_{t}\in [0,1]$ is the (number-weighted) fraction of agglomerates at time step $t\in T$, and $f_{\abbreviationPP},f_{\abbreviationAGG}\colon\R^2\to [0,\infty)$ denote the bivariate probability densities of the descriptor vector $(d(P),a(P))$ for primary particles and agglomerates, respectively, which do not depend on $t\in T$.

To achieve this goal, 
we develop an iterative approach, where the bivariate density $f$ determined in Section~\ref{sec.cop.mod} serves as initial choice for $f_{\abbreviationPP}$. It will be denoted by $f_{\abbreviationPP}^{(0)}$ in the following.
Furthermore, to get an initial choice for the bivariate density $f_{\abbreviationAGG}$ of agglomerates, denoted by $f_{\abbreviationAGG}^{(0)}$ from now on,
we  consider a  dataset of values  for the descriptor vector $(d(P),a(P))$ observed at the end of the experiment under consideration, and split it  into two subsets. The reason for this is that we cannot simply assume that the particle system observed at the end of an experiment consists almost exclusively of agglomerates.
Note that such a split was not necessary when determining the density 
$f_{\abbreviationPP}^{(0)}$ ( $=f$), 
since the dataset
$D_1\subset[0,\infty)\times[0,1]$ considered in Section~\ref{sec:FittingModelParams}, i.e., the set
 of all values $x=(x_1,x_2)$ available for   $(d(P),a(P))$ at time step $t=1$, mainly consists of data pairs $(x_1,x_2)$ corresponding to primary particles. 

To reliably determine a parametric (copula-based) model for the bivariate density $f_{\abbreviationAGG}^{(0)}$, we need a sufficiently large dataset with a high proportion of data pairs $(x_1,x_2)$ corresponding to agglomerates. 
Therefore, we consider the union $D_{\rm end}=\bigcup_{t=t_{\abbreviationEND}}^{52} D_t$  of datasets associated with several  time steps  close to the end of an experiment, where $t_{\abbreviationEND}=\min\{t\in T: n_t\le 1.05\, n_{52}\}$ and $D_t$ denotes the set
 of all values $x=(x_1,x_2)$ available for   $(d(P),a(P))$ at time step $t\in T$. For experiment DV1 it turned out that $t_{\abbreviationEND}=46$, which results in a number of particle descriptor vectors $|D_{\rm end}|=6128$ in $D_\abbreviationEND$, where the so-called cardinality $|A|$ of a set $A$ denotes the number of elements in $A$, i.e., in our case, the number of particle descriptor vectors. In contrast, if $A$ is a scalar, $|A|$ more commonly denotes the absolute value of $A$. As a consequence of $D_\abbreviationEND$ being the union of multiple sets $D_t$  its cardinality $|D_{\rm end}|$ is much larger than the number of particles $|D_{52}|=853$ observed at the last time step $t=52$ of experiment $\DVa$ as desired.

\begin{figure}[h]
    \centering
    \includegraphics[width=0.45\textwidth]{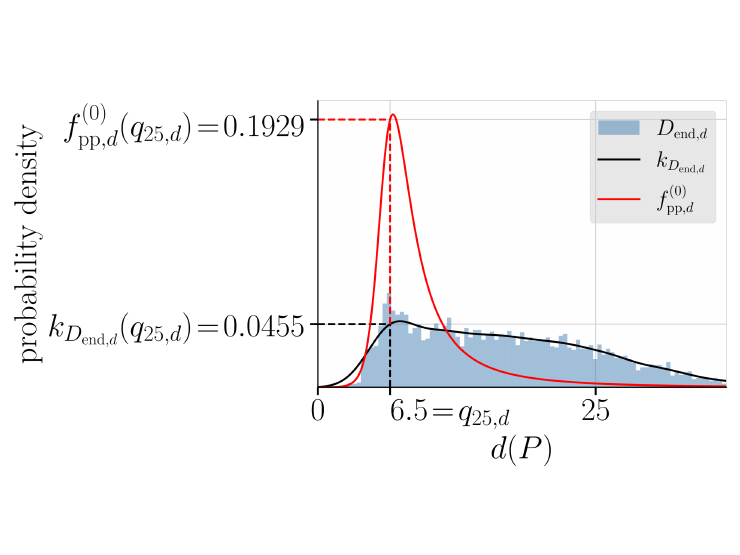}
    \caption{The values of $q_{25,d},\kernelDensityEstimation_{D_{\abbreviationEND, d}}(q_{25,d})$ and $f_{\abbreviationPP,d}^{(0)}(q_{25,d})$, used for computing the estimate $\widehat{\MixingRatio}_\abbreviationEND$
    given in Eq.~(\ref{eq:MixingRatioQuantile}).}
    \label{fig:est_m_end}
\end{figure}

Moreover,  to obtain a subset of $D_{\rm end}=\{(x_1^{(1)},x_2^{(1)}),\ldots,(x_1^{(|D_{\rm end}|)},x_2^{(|D_{\rm end}|)})\}$, which mainly consists of particle descriptor vectors corresponding to agglomerates, 
we determine an estimate for the (number-weighted) fraction
 $\MixingRatio_{\ExpEnd}\in [0,1]$ 
of agglomerates observed in $D_{\rm end}$, see Figure \ref{fig:est_m_end}.

To get a formula for such an estimate, we assume  that the dataset $D_{\rm end}$ is a sample drawn from the bivariate density
$f_\abbreviationEND\colon \mathbb{R}^2\rightarrow [0,\infty)$ which  has the following form:
\begin{equation}\label{eq:f_mix_end}
    f_{\abbreviationEND}(x)=(1-\MixingRatio_{\ExpEnd})f_{\abbreviationPP}(x)+\MixingRatio_{\ExpEnd} f_{\abbreviationAGG}(x)\qquad\mbox{for each $x\in\R^2$,}
\end{equation}
where $f_{\abbreviationPP}$ and $f_{\abbreviationAGG}$ are the same bivariate probability densities as in Eq.~(\ref{eq:f_mix}). Furthermore, we assume that the minimum of the area-equivalent diameters of all  agglomerates considered  in the given experiment is larger than the 25 percent quantile  $q_{25, d}$ of the marginal density $f_d$ representing the area-equivalent diameter in $f_{\abbreviationPP}^{(0)}$, which has been fitted in Section~\ref{sec.cop.mod}.  This means that  $f_{\abbreviationAGG, d}(q_{25,d})=0$ and, consequently, Eq.~(\ref{eq:f_mix_end}) gives that
\begin{equation}\label{for.em.end}
    \MixingRatio_\ExpEnd=1-\frac{f_{\ExpEnd,d}(q_{25,d})}{f_{\abbreviationPP,d}(q_{25,d})},
\end{equation}
where $f_{\ExpEnd,d}$, $f_{\abbreviationPP,d}$ and $f_{\abbreviationAGG, d}$ denote the marginal densities 
representing the area-equivalent diameter in 
$f_{\ExpEnd}$, $f_{\abbreviationPP}$ and $f_{\abbreviationAGG}$, respectively.  Thus, substituting the density $f_{\ExpEnd,d}$ on the right-hand side of Eq.~(\ref{for.em.end}) 
by a Gaussian kernel density estimate
 $\kernelDensityEstimation_{D_{\abbreviationEND,d}}\colon \mathbb{R}\rightarrow [0, \infty)$ with a bandwidth obtained by Scott's rule \cite{scott2015}
 based on the dataset 
 $D_{\rm end, d}=\{x_1^{(1)},\ldots,x_1^{(|D_{\rm end}|)}\}$, and replacing $f_{\abbreviationPP,d}$ by 
$f_{\abbreviationPP,d}^{(0)}$,  we get the estimate \begin{equation} \label{eq:MixingRatioQuantile}
\widehat{\MixingRatio}_\abbreviationEND=1-\frac{\kernelDensityEstimation_{D_{\abbreviationEND,d}}(q_{25,d})}{f_{\abbreviationPP,d}^{(0)}(q_{25,d})}.
\end{equation}
Note that for the data obtained in experiment $\DVa$, it turned out that $ \widehat{\MixingRatio}_\abbreviationEND=0.764$, see Figure \ref{fig:est_m_end}.

As mentioned above, we can not assume that $D_{\rm end}$ consists exclusively of data corresponding to agglomerates. Thus, a fraction of $1-\widehat{\MixingRatio}_\abbreviationEND$ of data corresponding to primary particles, which are encompassed within the set $D_{\rm end}$, will be disregarded in order to fit the  initial choice $f_{\abbreviationAGG}^{(0)}$   for $f_{\abbreviationAGG}$ to the remaining data. To be precise, we use the bivariate density $f_{\abbreviationPP}^{(0)}$  to remove
\begin{equation}
    \delete=\lceil (1-\widehat{\MixingRatio}_\abbreviationEND) |D_{\rm end}|\rceil
\end{equation}
descriptor vectors  from $D_{\rm end}$ that are likely to correspond to primary particles, where $\lceil x\rceil$ denotes the smallest integer which is larger than or equal to $x>0$, and the remaining $|D_{\rm end}|-\delete$ data vectors are considered to correspond to agglomerates.

First, we draw a sample $ \deleteSet=\{(d_i,a_i)\colon i=1,\dots,\delta\}\subset[0,\infty)\times[0,1]$ 
of $\delete$ particle descriptor vectors 
from a (slightly) truncated version of $f_{\abbreviationPP}^{(0)}$, using the accept-reject method \cite{Robert2004}. 
Note that this method facilitates sampling a so-called target density $f_\mathrm{tar}:\R^2\to[0,\infty)$, by sampling a so-called instrumental density $f_\mathrm{ins}:\R^2\to[0,\infty)$ such that  $f_\mathrm{tar}$ and $f_\mathrm{ins}$  must satisfy the inequality  $f_\mathrm{tar}(x)\leq M\, f_\mathrm{ins}(x)$ for each $x\in\{y\in\R^2: f_\mathrm{tar}(y)>0\}$  of the support of $f_\mathrm{tar}$, where $M>0$ is 
some constant. 

In our case, $f_\mathrm{tar}$ is given by a truncation  of $f_{\abbreviationPP}^{(0)}$ whose support is restricted to the interval $[0,q_{99.9}]\times [0,1]$. Here,  $q_{99.9}<\infty$ denotes the 99.9 percent quantile of the marginal probability density $f_{\abbreviationPP, d}^{(0)}$ of $f_{\abbreviationPP}^{(0)}$,
which represents the area-equivalent diameter of primary particles. Using this truncated version of $f_{\abbreviationPP}^{(0)}$, which is 
is a bounded function with bounded support,
rules out  values of   area-equivalent diameter and aspect ratio  outside their physically plausible ranges. 
Furthermore,   for the instrumental density  $f_\mathrm{ins}$,
we  select the probability density of  the uniform distribution  on the rectangle $[0,q_{99.9}]\times [0,1]$, putting $M=q_{99,9}\max\{f_{\abbreviationPP}^{(0)}(x), x\in [0,q_{99.9}]\times [0,1]\}$. Then, we can sample $f_\mathrm{tar}$ by sampling $f_\mathrm{ins}$ in the following way: We draw a realization $x\in [0,q_{99.9})\times [0,1]$ from $f_\mathrm{ins}$ and a realization $u\in [0,1]$ from the  uniform distribution on the interval $[0,1]$. If  $u\leq f_\mathrm{tar}(x)/(M\, f_\mathrm{ins}(x))$, we ``acccept'' $x$  and, otherwise, we ``reject'' it. We repeat this procedure until we get ``accepted'' a total of $\delete$ sample values. 

The dataset $ \deleteSet=\{(d_i,a_i)\colon i=1,\dots,\delta\}\subset[0,\infty)\times[0,1]$ obtained in this way
is now used to remove $\delta$ descriptor vectors from $D_{\rm end}$. 
Since the values of  area-equivalent diameter and aspect ratio are on different scales, we have to normalize these quantities. For this we use the empirical standard deviations
\begin{equation}\label{def.sta.dev}
\sigma_1=\sqrt{\frac{1}{|D_{\rm end}|-1}\sum_{i=1}^{|D_{\rm end}|}(x_1^{(i)}-\mu_1)^2},\qquad
\sigma_2=\sqrt{\frac{1}{|D_{\rm end}|-1}\sum_{i=1}^{|D_{\rm end}|}(x_2^{(i)}-\mu_2)^2},
\end{equation}
where 
$\mu_1=|D_{\rm end}|^{-1}\sum_{(\areaEquivalentDiameter,\aspectRatio)\in D_{\rm end}} d  $ and $\mu_2=|D_{\rm end}|^{-1}\sum_{(\areaEquivalentDiameter,\aspectRatio)\in D_{\rm end}} a$. In the next step, we  determine an injective function $\assignment\colon\{1,\dots,\delete\}\to D_{\rm end}$ such that the descriptor vectors 
$(x_1^{(p(1))},x_2^{(p(1))}),\ldots,(x_1^{(p(\delta))},x_2^{(p(\delta))}) \in D_{\rm end}$ can be considered to correspond to primary particles and, thus, be removed from $D_{\rm end}$. 
To find a suitable function $\assignment$, we 
solve the following optimization problem: 
\begin{equation}\label{eq:LinearAssignmentProblem}
\min_\assignment\sum_{i=1}^\delete \Biggl\lVert\bigg(\frac{d_i-x_1^{(p(i))}}{\sigma_1},\,\frac{a_i-x_2^{(p(i))}}{\sigma_2}\bigg)\Biggr\rVert
\end{equation}
where $\lVert\cdot\rVert$ denotes the Euclidean norm in $\R^2$.
Note that the minimization problem given in Eq.~(\ref{eq:LinearAssignmentProblem}) can be easily reformulated as a linear assignment problem \cite{Karp1980}. Its solution, denoted by  $\assignment^\StarSymbol\colon\{1,\dots,\delete\}\to D_{\rm end}$, will be used to remove the descriptor vectors
$(x_1^{(p^\StarSymbol(1))},x_2^{(p^\StarSymbol(1))}),\ldots,(x_1^{(p^\StarSymbol(\delta))},x_2^{(p^\StarSymbol(\delta))})$ from $D_{\rm end}$, which leads to the reduced dataset
\begin{equation} \label{red.dat.set}
D_{\abbreviationAGG}=D_{\rm end}\setminus\Bigl\{(x_1^{(p^\StarSymbol
(1))},x_2^{(p^\StarSymbol(1))}),\ldots,(x_1^{(p^\StarSymbol(\delta))},x_2^{(p^\StarSymbol(\delta))})\Bigr\}.
\end{equation}
Finally, using the methods stated in Section~\ref{sec.cop.mod},  we determine the bivariate probability density
$f_{\abbreviationAGG}^{(0)}$ by fitting it to the dataset $D_{\abbreviationAGG}$ given in Eq.~(\ref{red.dat.set}).

\subsection{Iterative adjustment of the  bivariate probability densities $f_{\abbreviationPP}^{(0)}$ and $f_{\abbreviationAGG}^{(0)}$}\label{sec:iterative}

In Sections~\ref{sec.cop.mod}
and 
\ref{sec:FittingModelParams} we showed how  the initial fits $f_{\abbreviationPP}^{(0)}$ and $f_{\abbreviationAGG}^{(0)}$ can be determined for the bivariate probability densities  $f_{\abbreviationPP}$ and $f_{\abbreviationAGG}$ of particle descriptor vectors of primary particles and agglomerates, respectively. 
However, these fits are based on the assumption that the dataset $D_1$, introduced in Section~\ref{sec.cop.mod}, consists almost exclusively of desciptor vectors corresponding to primary particles.   Thus, possible agglomerates encompassed within this dataset  are erroneously considered for fitting $f_{\abbreviationPP}^{(0)}$. Furthermore, since we used $f_{\abbreviationPP}^{(0)}$ to obtain the dataset $D_{\abbreviationAGG}$ given in Eq.~(\ref{red.dat.set}),  our assumption made on $D_1$ also affects the fit of $f_{\abbreviationAGG}^{(0)}$.

We now describe a procedure for adjusting $f_{\abbreviationPP}^{(0)}$ and $f_{\abbreviationAGG}^{(0)}$ iteratively with the goal to successively improve
their goodness-of-fit, which results in a sequence of pairs of probability densities $(f_{\abbreviationPP}^{(\iteration)}, f_{\abbreviationAGG}^{(\iteration)})$, for $\iteration=1,\dots,\NumIterations$ with $\NumIterations\ge 2$ being the number of iteration steps in which we each adjust exactly one of the two probability densities. More precisely, in each odd iteration step $\iteration\in\{1,\dots,\NumIterations\}$, we determine a new   probability density  $f_{\abbreviationPP}^{(\iteration)}$  of primary particles, leaving the probability density  $f_{\abbreviationAGG}^{(\iteration-1)}$ of agglomerates unchanged. Vice versa, in each even iteration step  $\iteration\in\{1,\dots,\NumIterations\}$, we determine a new  probability density  $f_{\abbreviationAGG}^{(\iteration)}$ of agglomerates and leave the probability density  $f_{\abbreviationPP}^{(\iteration-1)}$  of primary particles unchanged. Pseudocode for the iterative fitting procedure is provided in Algorithm~\ref{alg:iter_adjustments}.

\begin{algorithm}[h]
	\DontPrintSemicolon
	\caption{Iterative adjustments and selection of best fits}\label{alg:iter_adjustments}
	\KwIn{initially fitted densities $f_{\abbreviationPP}^{(0)}$ and $f_{\abbreviationAGG}^{(0)}$, sets of particle descriptor vectors $D_1$ and $D_\abbreviationEND$.}
	\KwOut{final fits $f_{\abbreviationPP}^{\StarSymbol}$ and $f_{\abbreviationAGG}^{\StarSymbol}$} 
	$\cdot$ Compute $\widehat{m}_1^{(1)}$ using Eq. (22).\;
	$\cdot$ Create $D_\mathrm{pp}^{(1)}$ using Eqs. (24) and (25).\;
	$\cdot$ Fit $f_\mathrm{pp}^{(1)}$ to $D_\mathrm{pp}^{(1)}$ and put $f_\mathrm{agg}^{(1)}$ equal to $f_\mathrm{agg}^{(0)}$.\;
	\For{$j\in\{2,\ldots, 50\}$}{
		\For{$z\in\{1,\mathrm{end}\}$}{
			$\cdot$ Compute $\widehat{m}_z^{(j)}$ and $l_z^{(j)}$ using Eqs. (28) and (31).\;
		}
		\If{$j\geq 3$}{
			\If{$|\widehat{\MixingRatio}_1^{(\iteration)}-\widehat{\MixingRatio}_1^{(\iteration-1)}|<0.001$\textnormal{ and } $|\widehat{\MixingRatio}_\abbreviationEND^{(\iteration)}-\widehat{\MixingRatio}_\abbreviationEND^{(\iteration-1)}|<0.001$}{
				$\cdot$ $J = j$\;
				\textbf{break}\;
			}
		}
		\If{$\iteration$\textnormal{ odd}}{
			$\cdot$ Compute $\mathrm{pp}_1(x)$ for all $x\in D_1$ using Eq. (30).\;
			$\cdot$ Create $D_\mathrm{pp}^{(j)}$.\;
			$\cdot$ Fit $f_\mathrm{pp}^{(j)}$ to $D_\mathrm{pp}^{(j)}$ and put $f_\mathrm{agg}^{(j)}$ equal to $f_\mathrm{agg}^{(j-1)}$.\;
		}
		\ElseIf{$\iteration$\textnormal{ even}}{
			$\cdot$ Compute $\mathrm{pp}_\mathrm{end}(x)$ for all $x\in D_\mathrm{end}$ using Eq. (30).\;
			$\cdot$ Create $D_\mathrm{agg}^{(j)}$.\;
			$\cdot$ Fit $f_\mathrm{agg}^{(j)}$ to $D_\mathrm{agg}^{(j)}$ and put $f_\mathrm{pp}^{(j)}$ equal to $f_\mathrm{pp}^{(j-1)}$.\;
		}
	}
	$\cdot$ Determine function $\rho$ using Eq. (32).\;
	$\cdot$ $\displaystyle\iteration^\StarSymbol=\argmin_{\iteration\in\{2,\ldots,\NumIterations\}}\rho(\iteration)-1$\;
	$\cdot$ $f_\mathrm{pp}^\StarSymbol=f_\mathrm{pp}^{(\iteration^\StarSymbol)},\,f_\mathrm{agg}^\StarSymbol=f_\mathrm{agg}^{(\iteration^\StarSymbol)}$\;
\end{algorithm}

\subsubsection{Algorithm for the computation of $f_{\abbreviationPP}^{(1)}$}\label{sec:comp_f_pp_1}
\begin{figure}[h]
    \centering
    \includegraphics[width=0.45\textwidth]{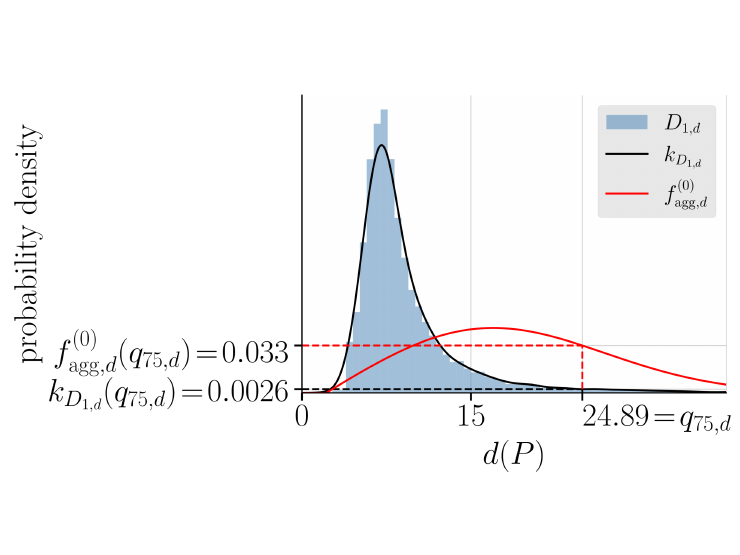}
    \caption{The values of $q_{75,d},\kernelDensityEstimation_{D_{1, d}}(q_{75,d})$ and $f_{\abbreviationAGG,d}^{(0)}(q_{75,d})$, used for computing the estimate $\widehat{\MixingRatio}_1^{(1)}$ given
    in Eq.~(\ref{eq:mixing_ratio_estimate_iter1}).}
    \label{fig:est_m_1}
\end{figure}

As stated in Section~\ref{sec.cop.mod}, the probability density $f_{\abbreviationPP}^{(0)}$ was fitted to the complete set $D_1$. In the fist iteration step $\iteration=1$ we  now replace  $f_{\abbreviationPP}^{(0)}$  by $f_{\abbreviationPP}^{(1)}$, where we fit $f_{\abbreviationPP}^{(1)}$
to a subset $D^{(1)}_{\abbreviationPP}\subseteq D_1$ which ideally consists of the particle descriptor vectors of all primary particles observed at time $t=1$, excluding those of agglomerates. 
For this,  we first determine an estimate of the  fraction ${\MixingRatio}_1\in [0,1]$ of agglomerates in $D_1$.  Similar to Section~\ref{sec:FittingModelParams}, where we determined a 25 percent quantile, we now consider the 75 percent quantile $q_{75,d}$ of the marginal density of $f_{\abbreviationAGG,d}^{(0)}$, which represents the area-equivalent diameters of agglomerates, and assume that  $q_{75,d}$ is larger than the maximum of the area-equivalent diameters of all primary particles considered in the given experiment. This means that 
 $f_{\abbreviationPP,\areaEquivalentDiameter}(q_{75,d})=0$ and, consequently,   Eq.~(\ref{eq:f_mix}) gives that
\begin{equation}\label{em.onefor}
\MixingRatio_1=\frac{f_{1,d}(q_{75,d})}{f_{\abbreviationAGG,d}(q_{75,d})},
\end{equation}
where $f_{1,d}$ denotes the marginal density 
representing the area-equivalent diameter in 
$f_1$. Thus, like in Section~\ref{sec:FittingModelParams},
substituting the density $f_{1,d}$ on the right-hand side of Eq.~(\ref{em.onefor})
by a Gaussian kernel density estimate
 $\kernelDensityEstimation_{D_{1,d}}\colon \mathbb{R}\rightarrow [0, \infty)$ 
 based on the first components of the two-dimensional vectors in the
 dataset 
 $D_1$, and replacing $f_{\abbreviationAGG,d}$ by the marginal density
$f_{\abbreviationAGG,d}^{(0)}$ 
representing the area-equivalent diameter in $f_{\abbreviationAGG}^{(0)}$,  we get the estimate 
\begin{equation}\label{eq:mixing_ratio_estimate_iter1}
    \widehat{\MixingRatio}_1^{(1)}=\frac{\kernelDensityEstimation_{D_{1,d}}(q_{75,d})}{f_{\abbreviationAGG,d}^{(0)}(q_{75,d})},
\end{equation}
where the superscript in  $\widehat{\MixingRatio}_1^{(1)}$ indicates the number  of the current (first) iteration step.
Note that for the data obtained in experiment DV1, it turned out that $ \widehat{\MixingRatio}_1^{(1)}=0.079$, see Figure \ref{fig:est_m_1}.

We now show how the estimate $\widehat{\MixingRatio}_1^{(1)}$ given in Eq.~(\ref{eq:mixing_ratio_estimate_iter1})  can be used for getting a subset $D^{(1)}_{\abbreviationPP}\subseteq D_1$. 
In principle, we could proceed as in Section~\ref{sec:FittingModelParams},  where we determined the subset $D_\abbreviationAGG\subseteq D_{\rm end}$  using the initial fit  $f_{\abbreviationPP}^{(0)}$ of  $f_{\abbreviationPP}$. Analogously to this, we could exploit  the initial fit  $f_{\abbreviationAGG}^{(0)}$ of  $f_{\abbreviationAGG}$,
derived in Section~\ref{sec:FittingModelParams}, in order to get  a suitable subset $D^{(1)}_{\abbreviationPP}\subseteq D_1$. Instead, we will use  another approach that includes both initial fits $f_{\abbreviationPP}^{(0)}$ and $f_{\abbreviationAGG}^{(0)}$ of $f_{\abbreviationPP}$ and $f_{\abbreviationAGG}$, respectively. Using fits of the probability densities
$f_{\abbreviationPP}$ and $f_{\abbreviationAGG}$
of both particle types, primary particles and agglomerates, we expect to get a subset $D^{(1)}_{\abbreviationPP}\subseteq D_1$
that is more representative for particle descriptor vectors of primary particles than just using  a variant of the approach stated in Section~\ref{sec:FittingModelParams}.

More precisely, we consider the estimate $f_1^{(1)}\colon\R^2\rightarrow [0, \infty)$  of the bivariate probability density $f_1$ given in Eq.~(\ref{eq:f_mix}), where 
\begin{equation}\label{eq:probability_iteration_1}
    f_1^{(1)}(x)=(1-\widehat{\MixingRatio}_1^{(1)})f_{\abbreviationPP}^{(0)}(x)+\widehat{\MixingRatio}_1^{(1)} f_{\abbreviationAGG}^{(0)}(x)\qquad\mbox{for each $x\in\R^2$.}
\end{equation}
The estimate $f_1^{(1)}$ of $f_1$ given in Eq.~(\ref{eq:probability_iteration_1})
will be used to determine the estimate 
\begin{equation}\label{eq:class1}
    \abbreviationPP^{(1)}(x)=(1-\widehat{\MixingRatio}_1^{(1)})\frac{f_{\abbreviationPP}^{(0)}(x)}{f_1^{(1)}(x)}
\end{equation}
of the conditional probability 
$ \abbreviationPP(x)=(1-\MixingRatio_1)f_{\abbreviationPP}(x)/f_1(x)$
that a particle observed at time step $t=1$ is a primary particle, under the condition that this particle has the descriptor vector $x$, for any given $x\in[0,\infty)\times[0,1]$. This leads to the  subset  
\begin{equation}\label{eq:agg_subset.pp}
    D^{(1)}_{\abbreviationPP}=\Big\{x\in D_1\colon \abbreviationPP^{(1)}(x)\geq U^{(1)}_x\Big\}
\end{equation}
of $D_1$,
where $\{U^{(1)}_x, x\in D_1\}$ is a family of 
independent  random variables, which are uniformly distributed on the interval $[0,1]$.  Thus, by means of the methods stated in Section~\ref{sec.cop.mod},  we can determine the bivariate probability density
$f_{\abbreviationPP}^{(1)}$ by fitting it to the dataset $D^{(1)}_{\abbreviationPP}$ given in Eq.~(\ref{eq:agg_subset.pp}). Afterwards, $f_{\abbreviationAGG}^{(1)}$ is put equal to $f_{\abbreviationAGG}^{(0)}$ as indicated in the beginning of  Section~\ref{sec:iterative}, where we stated that we only adjust one of the two densities in each iteration step.

\subsubsection{Procedure for subsequent adjustment of  $f_{\abbreviationPP}^{(1)}$ and $f_{\abbreviationAGG}^{(1)}$}\label{sec:procedure_subseq_adj}

 Recall that  we fitted the bivariate probability densities 
$f_{\abbreviationPP}^{(1)}$ and $f_{\abbreviationAGG}^{(1)}$ ($=f_{\abbreviationAGG}^{(0)}$)  to suitably chosen subsets of $D_1$ and $D_\ExpEnd$, respectively. For constructing these subsets, we estimated the fractions $m_1$ and
 $m_{\rm end}$  
 of agglomerates in $D_1$ and $D_\ExpEnd$  by considering the  25  and 75 percent quantiles $q_{25,d}$ and $q_{75,d}$, where we assumed that $f_{\abbreviationAGG,d}(q_{25,d})= f_{\abbreviationPP,d}(q_{75,d})=0$. From now on,  for each iteration step $\iteration\in\{2,\ldots,\NumIterations\}$,
  we consider a different approach in order to determine  estimates $\widehat{\MixingRatio}^{(\iteration)}_1$ and $\widehat{\MixingRatio}^{(\iteration)}_\abbreviationEND$ of  $m_1$ and $m_{\abbreviationEND}$, respectively, 
  which will be used for
  constructing subsets $D_{\abbreviationPP}^{(\iteration)}\subseteq D_1$ and $D_\abbreviationAGG^{(\iteration)}\subseteq D_\abbreviationEND$,
  to successively improve the fits for $f_{\abbreviationPP}$ and $f_{\abbreviationAGG}$.  
  As indicated at the beginning of Section \ref{sec:iterative},  we then fit $f_{\abbreviationPP}^{(\iteration)}$ to $D_{\abbreviationPP}^{(\iteration)}$ if $\iteration$ is odd, and fit $f_\abbreviationAGG^{(\iteration)}$ to $D_\abbreviationAGG^{(\iteration)}$ if $\iteration$ is even. 
  We iterate this procedure for iteratively increasing iteration steps  until the following termination condition is met for some iteration step $j'\geq 3$: if $|\widehat{\MixingRatio}_1^{(\iteration')}-\widehat{\MixingRatio}_1^{(\iteration'-1)}|<0.001$ and $|\widehat{\MixingRatio}_\abbreviationEND^{(\iteration')}-\widehat{\MixingRatio}_\abbreviationEND^{(\iteration'-1)}|<0.001$, i.e., the estimates for $\MixingRatio_1$ and $\MixingRatio_\abbreviationEND$ have not changed much compared to the estimates from the previous iteration, we stop iterating. In order to guarantee termination of this procedure, the procedure automatically is stopped after iteration step 50. Thus, the number of iterations $J$ is given by
   \begin{equation}\label{eq:NumIterations}
      \NumIterations=\min\{\iteration',50\}. 
  \end{equation}
  Note that, in our experiments this condition was never met, i.e., the number of iterations $J$ was always below 50. More details on the convergence of the iterative fitting procedure can be found in Section~\ref{sec:estimates} and in Figure~\ref{fig:mixing_ratios_and_rank_per_iteration}.
  
   To facilitate the explanation of the following steps, we introduce the variable $\BeginOrEnd\in\{1,\abbreviationEND\}$. Specifically, for any $\iteration\in\{2,\ldots,\NumIterations\}$ and $\BeginOrEnd\in\{1,\abbreviationEND\}$, we consider the likelihood function
\begin{equation}\label{eq:LikelihoodW}
\mathcal{L}_\BeginOrEnd^{(\iteration)}(\MixingRatio ; D_\BeginOrEnd)=\prod_{x\in D_\BeginOrEnd}\left((1-\MixingRatio)f_{\abbreviationPP}^{(\iteration-1)}(x)+\MixingRatio f_{\abbreviationAGG}^{(\iteration -1)}(x)\right)
\end{equation}
with some fraction $m\in [0,1]$ of agglomerates. The (maximum likelihood) estimate $\widehat{\MixingRatio}_{\BeginOrEnd}^{(\iteration)}$ for the fraction of agglomerates in $D_\BeginOrEnd$ is then given by
\begin{equation}\label{eq:MixingRatioML}  \widehat{\MixingRatio}_{\BeginOrEnd}^{(\iteration)}=\argmax_{\MixingRatio\in [0,1]}\mathcal{L}_{\BeginOrEnd}^{(\iteration)}(\MixingRatio ; D_\BeginOrEnd).
\end{equation}

Note that in the definitions of the estimators 
$
\widehat{\MixingRatio}_\abbreviationEND^{(1)} (=
\widehat{\MixingRatio}_\abbreviationEND)$
and
$\widehat{\MixingRatio}_1^{(1)}$
given in Eqs.~(\ref{eq:MixingRatioQuantile}) and (\ref{eq:mixing_ratio_estimate_iter1}) 
only  the area-equivalent diameters of primary particles and agglomerates are taken into account, but not their aspect ratios. 
However, after having determined  the
 bivariate probability densities 
$f_{\abbreviationPP}^{(1)}$ and $f_{\abbreviationAGG}^{(1)}$ in the first step, we  now use the recursive definition of the likelihood function given in Eq.~\eqref{eq:LikelihoodW} in order to compute
 the estimate $\widehat{\MixingRatio}_{\BeginOrEnd}^{(\iteration)}$  based on descriptor vectors $x \in D_\BeginOrEnd$ for any $\iteration\in\{2,\ldots,\NumIterations\}$ and $\BeginOrEnd\in\{1,\abbreviationEND\}$, i.e., in this manner we determine  an 
estimate $\widehat{\MixingRatio}_{\BeginOrEnd}^{(\iteration)}$ of the fraction 
 $m_\BeginOrEnd$
 of agglomerates,
 which provides a good fit with respect to both the  area-equivalent diameter  and  aspect ratio of particles.

Recall that in Eq.~\eqref{eq:probability_iteration_1}  we estimated   $f_1$ by $f_1^{(1)}$ using the previously determined  estimates $f_\abbreviationPP^{(0)}, f_\abbreviationAGG^{(0)}$ and $\widehat{\MixingRatio}_1^{(1)}$. 
 For each $j\in\{2,\ldots, \NumIterations\}$,
in order to further improve the fit of the bivariate probability densities $f_1$ and $f_{\rm end}$ given in
Eqs.~(\ref{eq:f_mix}) and (\ref{eq:f_mix_end}), 
we now use  the estimates $f_{\abbreviationPP}^{(\iteration-1)},f_{\abbreviationAGG}^{(\iteration-1)}$
of 
$f_{\abbreviationPP},f_{\abbreviationAGG}$
obtained in the previous iteration step, as well as the estimate $\widehat{\MixingRatio}_\BeginOrEnd^{(\iteration)}$ given in Eq.~\eqref{eq:MixingRatioML}  for $\BeginOrEnd\in\{1,\abbreviationEND\}$. More precisely, for $\BeginOrEnd\in\{1,\abbreviationEND\}$, 
we estimate $f_\BeginOrEnd$ by $f^{(\iteration)}_\BeginOrEnd\colon \R^2\rightarrow [0,\infty)$, where   
\begin{equation}\label{eq:probability_iteration_z}
    f_\BeginOrEnd^{(\iteration)}(x)=(1-\widehat{\MixingRatio}_\BeginOrEnd^{(\iteration)})f_{\abbreviationPP}^{(\iteration-1)}(x)+\widehat{\MixingRatio}_\BeginOrEnd^{(\iteration)} f_{\abbreviationAGG}^{(\iteration-1)}(x)\qquad\mbox{for each $x\in\R^2$.}
\end{equation}
Furthermore, similar to Eq.~(\ref{eq:class1}), the estimate $f_\BeginOrEnd^{(\iteration)}$ of $f_\BeginOrEnd$ given in Eq.~(\ref{eq:probability_iteration_z})
will be used to determine the estimate 
\begin{equation}\label{eq:class1_z}
    \abbreviationPP^{(\iteration)}_\BeginOrEnd(x)=(1-\widehat{\MixingRatio}_\BeginOrEnd^{(\iteration)})\frac{f_{\abbreviationPP}^{(\iteration -1)}(x)}{f_\BeginOrEnd^{(\iteration)}(x)}
\end{equation}
of the conditional probability 
$ \abbreviationPP_\BeginOrEnd(x)=(1-\MixingRatio_\BeginOrEnd)f_{\abbreviationPP}(x)/f_\BeginOrEnd(x)$
that a particle observed in $D_\BeginOrEnd$
is a primary particle, under the condition that it has the descriptor vector $x$, for any given $x\in[0,\infty)\times[0,1]$. This leads to the  subsets
$D^{(\iteration)}_{\abbreviationPP}=\big\{x\in D_1\colon \abbreviationPP_1^{(\iteration)}(x)\geq U^{(j)}_{1,x}\big\}$
and $D^{(\iteration)}_{\abbreviationAGG}=D_\abbreviationEND\setminus \big\{x\in D_\abbreviationEND\colon \abbreviationPP_\abbreviationEND^{(\iteration)}(x)\geq U^{(j)}_{\abbreviationEND,x}\big\}$
of $D_1$ and $D_\abbreviationEND$, respectively.
Here, $\{U^{(j)}_{1,x}, x\in D_1\}$ and $\{U^{(j)}_{\abbreviationEND,x}, x\in D_\abbreviationEND\}$ are families of 
independent  random variables, which are uniformly distributed  on the interval $[0,1]$. Finally, by means of the methods stated in Section~\ref{sec.cop.mod},  we  determine either
$f_{\abbreviationPP}^{(\iteration)}$ or $f_{\abbreviationAGG}^{(\iteration)}$, depending on whether $\iteration$ is odd or even,  fitting the bivariate  probability densities 
$f_{\abbreviationPP}^{(\iteration)}$ and $f_{\abbreviationAGG}^{(\iteration)}$ to the datasets $D^{(j)}_{\abbreviationPP}$ and  $D^{(j)}_{\abbreviationAGG}$, respectively.

\subsubsection{Determining the best fitting pair $(f_\abbreviationPP^\StarSymbol,f_\abbreviationAGG^\StarSymbol)$}
In the previous section we determined a sequence of fits $(f_\abbreviationPP^{(1)},f_\abbreviationAGG^{(1)}),\ldots,(f_\abbreviationPP^{(J)},f_\abbreviationAGG^{(J)}) $ for the bivariate probability densities  $f_{\abbreviationPP}$ and $f_{\abbreviationAGG}$  introduced in Eq.~\eqref{eq:f_mix}. However, in general, it is not clear if the goodness-of-fit of 
$(f_\abbreviationPP^{(j)},f_\abbreviationAGG^{(j)})$ improves with increasing $j$. 
Therefore, in the following, a rule is defined which measures the suitability of   $(f_\abbreviationPP^{(\iteration)},f_\abbreviationAGG^{(\iteration)})$. Recall that by means of Eqs.~\eqref{eq:LikelihoodW} and (\ref{eq:MixingRatioML}) we determined the maximum likelihood estimates $\widehat{\MixingRatio}^{(\iteration)}_1$ and $\widehat{\MixingRatio}^{(\iteration)}_{\rm end}$ 
for the fractions of agglomerates in  $D_1$ and $D_\abbreviationEND$, respectively, based on the knowledge of $f_\abbreviationPP^{(\iteration-1)}$ and $f_\abbreviationAGG^{(\iteration-1)}$. We now employ the likelihoods  
\begin{equation}
\evaluation_{\BeginOrEnd}^{(\iteration)}=\mathcal{L}_{\BeginOrEnd}^{(\iteration)}(\widehat{\MixingRatio}_{\BeginOrEnd}^{(\iteration)} ; D_\BeginOrEnd), \qquad\mbox{
where  $\iteration\in\{2,\ldots,\NumIterations\}$ and $\BeginOrEnd\in\{1,
\abbreviationEND\}$},
\end{equation}
as measures for the 
the suitability of   $(f_\abbreviationPP^{(\iteration)},f_\abbreviationAGG^{(\iteration)})$ for each $j\in\{1,\ldots,J-1\}$.  
More precisely, we consider the function $\rho\colon\{2,\ldots,\NumIterations\}\rightarrow\{2,\ldots,2(\NumIterations-1)\}$ 
which is given by
\begin{equation}
    \rho(\iteration)=\sum_{\BeginOrEnd\in\{1,\abbreviationEND\}}|\{\evaluation_\BeginOrEnd^{(i)}; i\in\{2,\ldots,\NumIterations\},\evaluation_\BeginOrEnd^{(i)}\geq\evaluation_\BeginOrEnd^{(\iteration)}\}|,
\end{equation}
for each iteration step $\iteration\in\{2,\ldots,\NumIterations\}$. Subsequently,  $\iteration^\StarSymbol=\argmin_{\iteration\in\{2,\ldots,\NumIterations\}}\rho(\iteration)-1\in\{1,\ldots,\NumIterations-1\}$ indicates the iteration step of our final fits $f_{\abbreviationPP}^\StarSymbol(=f_{\abbreviationPP}^{(\iteration^\StarSymbol)})$ and $f_{\abbreviationAGG}^\StarSymbol(=f_{\abbreviationAGG}^{(\iteration^\StarSymbol)})$ for $f_{\abbreviationPP}$ and $f_{\abbreviationAGG}$, respectively, where the corresponding sum $\rho(\iteration^\StarSymbol)$ of ranks of the likelihoods $\evaluation_1^{(\iteration^\StarSymbol+1)}$ and $\evaluation_\abbreviationEND^{(\iteration^\StarSymbol+1)}$ is the smallest. 

Although we fitted $f_{\abbreviationPP}^\StarSymbol$ and $f_{\abbreviationAGG}^\StarSymbol$ to the sets $D_1$ and $D_\abbreviationEND$, respectively, we assume from   now on  that $f_{\abbreviationPP}^\StarSymbol$ and $f_{\abbreviationAGG}^\StarSymbol$ are  well-suited for modeling the mixing components $f_{\abbreviationPP}$ and $f_{\abbreviationAGG}$ of the bivariate probability density $f_t$, given in Eq.~(\ref{eq:f_mix}), for each time step $t\in T$.

\subsubsection{Estimating the fraction of agglomerates  $\MixingRatio_{t}$  for all  $t\in T$}\label{sec:est_frac_agg}
\begin{figure}[h!]
    \centering
    \includegraphics[width=0.5\textwidth]{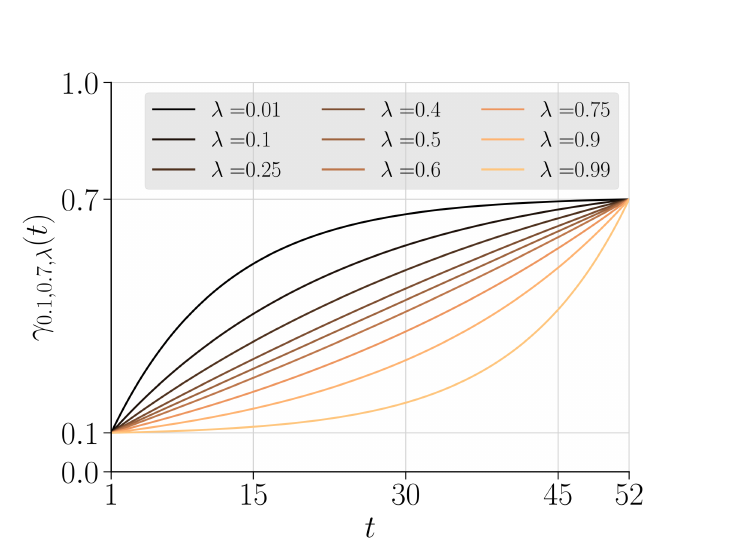}
    \caption{Visualization of $\pmrFunction_{\pmrMinimum,\pmrMaximum,\pmrContBerParam}$ with $\pmrMinimum=0.1$ and $\pmrMaximum=0.7$, for different values of $\pmrContBerParam\in (0,1)$.}
    \label{fig:cdf_continuous_bernoulli}
\end{figure}

After  determining the fits $f_{\abbreviationPP}^\StarSymbol$ and $f_{\abbreviationAGG}^\StarSymbol$ for the bivariate probability densities $f_{\abbreviationPP}$ and $f_{\abbreviationAGG}$ of  area-equivalent diameter and aspect ratio of primary particles and agglomerates, respectively, we can now estimate the fraction of agglomerates $\MixingRatio_{t}$  for each time step $t\in T$. For that purpose, we consider the  probability density 
$f_{m}^\StarSymbol = (1-m)f_{\abbreviationPP}^\StarSymbol+m f_{\abbreviationAGG}^\StarSymbol$ and the likelihood  
$  \mathcal{L}^\StarSymbol(m; D_t) = \prod_{x\in D_t}f_{m}^\StarSymbol(x)
$ for any $m\in[0,1]$ and $t\in T$. This leads to the maximum-likelihood estimate
\begin{equation}\label{eq:MixingRatioMaxLik}
    \widehat{\MixingRatio}_t=\argmax_{m\in [0,1]}\mathcal{L}^\StarSymbol(m; D_t)
\end{equation} 
of the fraction of agglomerates $m_t$ for each $t\in T$. Furthermore, we assume that the estimates $\{\widehat{\MixingRatio}_t, t\in T\}$ follow an underlying trend that is described by some parametric function   $\pmrFunction_{\pmrMinimum,\pmrMaximum,\pmrContBerParam}\colon T\rightarrow [\pmrMinimum,\pmrMaximum]$ such that
\begin{equation}\label{eq:pmr}
    \pmrFunction_{\pmrMinimum,\pmrMaximum,\pmrContBerParam}(t)=\pmrMinimum+(\pmrMaximum-\pmrMinimum)\pmrContBerFunction_\pmrContBerParam \left(\frac{t-1}{|T|-1}\right)\qquad\mbox{for each $t\in T$,}
\end{equation}
 i.e., we assume that $\widehat{\MixingRatio}_t\approx\pmrFunction_{\pmrMinimum,\pmrMaximum,\pmrContBerParam}(t)$ for each $t\in T$, with parameters  $\pmrMinimum\in[0,1],\pmrMaximum\in [\pmrMinimum,1]$ and $\pmrContBerParam\in (0,1)$. Note that the input $\frac{t-1}{|T|-1}=\frac{t-1}{52-1}$ of the function $\pmrContBerFunction_\pmrContBerParam\colon [0,1]\rightarrow [0,1]$ in Eq.~(\ref{eq:pmr}) is in $[0,1]$, for each $t\in T$, where $\pmrContBerFunction_\pmrContBerParam$ denotes the cumulative distribution function of the continuous Bernoulli distribution \cite{Loaiza2019}, which is given by
\begin{equation}
    \pmrContBerFunction_\pmrContBerParam(x)=\begin{cases}
    x,& \text{if } \pmrContBerParam=1/2,\\
    \displaystyle\frac{\pmrContBerParam^x(1-\pmrContBerParam)^{1-x}+\pmrContBerParam-1}{2\pmrContBerParam -1},              & \text{otherwise},
\end{cases}
\end{equation}
for each $x\in [0,1]$. 
Note that the  parameters $a,b,\lambda$ in Eq.~(\ref{eq:pmr}) are closely related with the  dynamics of the agglomeration experiments considered in this paper. Specifically, $\pmrMinimum$ and $\pmrMaximum$ represent the estimated fractions $\widehat{\MixingRatio}_1$ and $\widehat{\MixingRatio}_{52}$ of agglomerates at the beginning and end of the experiment, i.e.,  $\widehat{\MixingRatio}_1\approx\pmrFunction_{\pmrMinimum,\pmrMaximum,\pmrContBerParam}(1)=\pmrMinimum$ and $\widehat{\MixingRatio}_{52}\approx\pmrFunction_{\pmrMinimum,\pmrMaximum,\pmrContBerParam}(52)=\pmrMaximum$. Moreover, the function $\pmrFunction_{\pmrMinimum,\pmrMaximum,\pmrContBerParam}$
given in Eq.~(\ref{eq:pmr})
increases monotonously in $t\in T$, where $\pmrContBerParam<0.5$ indicates fast agglomeration at the beginning and slower agglomeration at the end of the experiment, whereas  $\pmrContBerParam>0.5$ indicates slow agglomeration at the beginning and faster agglomeration at the end of the experiment. The closer $\pmrContBerParam$ is to 0 or 1, the more apparent the respective behaviors become. For $\pmrContBerParam=0.5$, the fraction of agglomerates increases linearly, see 
Figure~\ref{fig:cdf_continuous_bernoulli}.

Finally, we determine the (optimal) parameter vector $ (\pmrMinimum^\StarSymbol,\pmrMaximum^\StarSymbol,\pmrContBerParam^\StarSymbol)\in[0,1]^2\times(0,1)$
by solving the non-linear least squares problem
\begin{equation}\label{eq:non-linear-least-squares}
    (\pmrMinimum^\StarSymbol,\pmrMaximum^\StarSymbol,\pmrContBerParam^\StarSymbol)=\argmin_{\pmrMinimum\in[0,1],\pmrMaximum\in [\pmrMinimum,1],\pmrContBerParam\in (0,1)}\sum_{t\in T}(\widehat{\MixingRatio}_t-\pmrFunction_{\pmrMinimum,\pmrMaximum,\pmrContBerParam}(t))^2.
\end{equation}
This gives  the estimate
\begin{equation}\label{eq:final_fit}
     f^\StarSymbol_{m_t^\StarSymbol}=(1-\MixingRatio_{t}^\StarSymbol)f_{\abbreviationPP}^\StarSymbol+\MixingRatio_{t}^\StarSymbol f_{\abbreviationAGG}^\StarSymbol,
\end{equation} 
of the bivariate probability density $f_t$
for each time step $t\in T$, where
$  \MixingRatio_{t}^\StarSymbol=\pmrFunction_{\pmrMinimum^\StarSymbol,\pmrMaximum^\StarSymbol,\pmrContBerParam^\StarSymbol}(t)$.

\subsection{Model validation}\label{sec:model_validation}

We now present some methods which we use to  check how well the probabilistic modeling approach developed in this paper fits the underlying image data. Recall that
in the previous sections we utilized the sets $D_1$ and $D_\abbreviationEND$ to fit the probability densities $f_\abbreviationPP^\StarSymbol$ and $f_\abbreviationAGG^\StarSymbol$ for primary particles and agglomerates, respectively. Since we used the same kind of primary particles in all five experiments $\DVa, \DVb, \UDVa,\UDVb$ and $\IBc$, a first aspect of model validation is to check
if the fitted densities $f_\abbreviationPP^\StarSymbol$ are similar  to each other for each of these experiments, see Section~\ref{sec:consistency_mixture_components}.
Furthermore,  for those  experiments  which utilized ultrasound treatment to disperse particles into almost exclusively primary particles, the fitted values of $ \pmrMinimum^\StarSymbol$
should be similar to each other, because they represent the (small) fractions of agglomerates, which are observed at the beginning of these experiments, see Section~\ref{sec:eval_a_b_lambda}. Finally, we  check if the probability densities $f^\StarSymbol_{m_t^\StarSymbol}$ fit sufficiently well  to their corresponding datasets $D_t$, for each $t\in T$, see Section~\ref{sec.fiv.thr}.
Note that for most time steps $t\in T$, the sets $D_t$ were not used for  fitting  the  components $f_\abbreviationPP^\StarSymbol$ and $f_\abbreviationAGG^\StarSymbol$ of the mixtures $f^\StarSymbol_{m_t^\StarSymbol}=(1-\MixingRatio_{t}^\StarSymbol)f_{\abbreviationPP}^\StarSymbol+\MixingRatio_{t}^\StarSymbol f_{\abbreviationAGG}^\StarSymbol$.

\subsubsection{The densities $f_\abbreviationPP^\StarSymbol$ and $f_\abbreviationAGG^\StarSymbol$}\label{sec:consistency_mixture_components}
Since the same type of particles is used in each of the five experiments $\DVa, \DVb, \UDVa,\UDVb$ and $\IBc$,  the  probability densities $f_\abbreviationPP^\StarSymbol$, which represent the joint distributions of area-equivalent diameter and aspect ratio  of primary particles, should be similar for all experiments listed in Table~\ref{tab:experiments}. On the other hand, the densities 
$f_\abbreviationAGG^\StarSymbol$ for 
agglomerates may  depend on the specificities of the respective experimental conditions as, for example, the used energy dissipation rate generated by different types of stirrers. However, similar densities $f_\abbreviationAGG^\StarSymbol$ should be fitted for $\DVa$ and $\DVb$ (as well as for  $\UDVa$ and $\UDVb$),  since these two experiments are repetitions of each other. It turned out that in all cases a quite good fit has been achieved, see Figure~\ref{fig:comparison_pp_and_agg} in Section~\ref{sec.res.ult}.

\subsubsection{The model parameters $\pmrMinimum^\StarSymbol, \pmrMaximum^\StarSymbol$ and $\pmrContBerParam^\StarSymbol$}\label{sec:eval_a_b_lambda}

In Section~\ref{sec:est_frac_agg}, see Eq.~(\ref{eq:pmr}), we introduced  the function $\pmrFunction_{\pmrMinimum,\pmrMaximum,\pmrContBerParam}:T\to[a,b]$ 
as a parametric approximation of the estimated fractions 
$\widehat{\MixingRatio}_t$ of agglomerates for each time step $t\in T$, where  we solved the non-linear least squares problem given in Eq.~(\ref{eq:non-linear-least-squares})
in order to  determine the optimal parameter values $\pmrMinimum^\StarSymbol,\pmrMaximum^\StarSymbol$ and $\pmrContBerParam^\StarSymbol$. To check 
whether the function $\pmrFunction_{\pmrMinimum,\pmrMaximum,\pmrContBerParam}$ 
given in Eq.~(\ref{eq:pmr})
is an appropriate choice for a parametric model, 
we will investigate the mean squared error $\mathrm{MSE}=\frac{1}{|T|}\sum_{t\in T}(\widehat{\MixingRatio}_t-\pmrFunction_{\pmrMinimum^\StarSymbol,\pmrMaximum^\StarSymbol,\pmrContBerParam^\StarSymbol}(t))^2$.

Similar to the situation explained in Section~\ref{sec:consistency_mixture_components}, where we highlighted some necessary consistencies between the probability densities $f_\abbreviationPP^\StarSymbol$ and $f_\abbreviationAGG^\StarSymbol$ for the different agglomeration experiments, some consistencies of $\pmrMinimum^\StarSymbol,\pmrMaximum^\StarSymbol$ and $\pmrContBerParam^\StarSymbol$ must be met. This is justified by the fact   that the same type of particles is used as feed in each of the five experiments $\DVa, \DVb, \UDVa,\UDVb$ and $\IBc$.
Moreover, we know how the particles are inserted into the stirring tank, where in addition to ethanol, ultrasound is used (except for $\UDVa$ and $\UDVb$) to further disperse the ``feed-suspension'', see Table~\ref{tab:experiments}.  Therefore, the fraction of agglomerates at the beginning of the experiments, which is represented by $\pmrMinimum^\StarSymbol$, should be similar for the experiments $\UDVa$ and $\UDVb$, as well as for  $\DVa, \DVb$ and $ \IBc$. Furthermore, the value of  $\pmrMinimum^\StarSymbol$ should be larger for $\UDVa$ and $\UDVb$ than for the other experiments, as there should be more agglomerates in the tank at the beginning of  $\UDVa$ and $\UDVb$. 

However, the similarities and dissimilarities of $\pmrMinimum^\StarSymbol$ mentioned above should be viewed with some caution as $\pmrMinimum^\StarSymbol$ represents  the fraction of agglomerates at time step $t=1$, which is not the actual beginning of an experiment. In fact, as stated in Section \ref{sec:acq_data}, the time step $t=1$ covers the period of 30-40 seconds after the ``feed-suspension'' is added to the tank. During this time period, agglomerates can already form and break, which can change the initial fraction of agglomerates. This is particularly relevant when comparing  $\pmrMinimum^\StarSymbol$ for experiments DV1, DV2, and IB3, as less agglomeration in the initial 30-40 seconds is expected for IB3 since it uses a lower energy dissipation rate compared to DV1 and DV2.

Lastly, we  check the similarity of the values of $\pmrMinimum^\StarSymbol, \pmrMaximum^\StarSymbol$ and $\pmrContBerParam^\StarSymbol$ which are obtained for the repeated experiments $\DVa$ and $\DVb$, as well as for $\UDVa$ and $\UDVb$.

\subsubsection{The probability densities $f^\StarSymbol_{m_t^\StarSymbol}$}\label{sec.fiv.thr}

 Finally, we  check if the probability densities $f^\StarSymbol_{m_t^\StarSymbol}$  introduced in Section \ref{sec:est_frac_agg}, see Eq.~(\ref{eq:final_fit}), fit sufficiently well  to their corresponding datasets $D_t$, for each $t\in T$.   To accomplish this we draw a sample of 10,000 realizations from  $f^\StarSymbol_{m_t^\StarSymbol}$, for each $t\in T$,  utilizing the accept-reject method. In particular, to get the  probability densities $f_\mathrm{tar}$ and $f_\mathrm{ins}$ required for this, we proceed in the same way as explained in Section~\ref{sec:FittingModelParams}. However,  we now use $f^\StarSymbol_{m_t^\StarSymbol}$ instead of $f_\abbreviationPP^{(0)}$ and, as a consequence,  the 99.9 percent quantile $q_{99.9}$ is determined for the marginal probability density $f^\StarSymbol_{m_t^\StarSymbol,d}$ of $f^\StarSymbol_{m_t^\StarSymbol}$, which represents the area-equivalent diameter of particles  at time step $t$. Subsequently,  for each time step $t\in T$, we  compute various characteristics, such as mean values,  standard deviations, correlation coefficients and quantiles, of the sampled two-dimensional  vector data and compare them with corresponding characteristics computed  for the dataset $D_t$.

\section{Results}\label{sec.res.ult}
In Section~\ref{sec:copula_approach} we presented the methods for  fitting    the mixture $f^\StarSymbol_{m_t^\StarSymbol}=(1-\MixingRatio_{t}^\StarSymbol)f_{\abbreviationPP}^\StarSymbol+\MixingRatio_{t}^\StarSymbol f_{\abbreviationAGG}^\StarSymbol$ 
to the
 dataset $D_t$, for each time step $t\in T$, where
we  illustrated the application of these methods by means of data from experiment $\DVa$.  We now systematically present the results which we obtained for all experiments listed in Table~\ref{tab:experiments}.

\subsection{Descriptive statistical analysis of measured image  data}

\begin{figure}[h]
    \centering
    \includegraphics[width=0.9\textwidth]{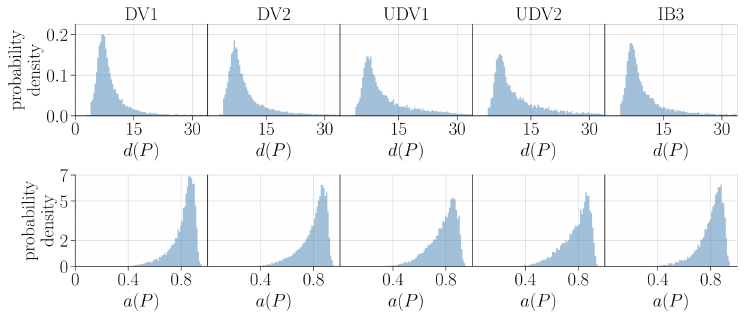}
    \caption{Histograms of area-equivalent diameter (top row) and aspect ratio (bottom row) of the descriptor vectors in $D_1$ of the experiments $\DVa$, $\DVb$, $\UDVa$, $\UDVb$ and $\IBc$ (from left to right).}
    \label{fig:t1_histograms}
\end{figure}

Recall that in Section~\ref{sec.cop.mod},  as an initial step for estimating $f_\abbreviationPP$, we fitted a bivariate probability density $f^{(0)}_\abbreviationPP$ to the complete set $D_1$, whereas we fitted $f^{(\iteration)}_\abbreviationPP$ to a subset of $D_1$, for each $\iteration\in\{1,\ldots,\NumIterations\}$. In fact, we initially assumed that $D_1$ consists exclusively of primary particles. This assumption was reasonable since the univariate distributions of both the area-equivalent diameter and the aspect ratio of particles measured at time step $t=1$ appear to be unimodal, as shown in Figure~\ref{fig:t1_histograms}. Note that this is also the case for experiments $\UDVa$ and $\UDVb$, where no additional ultrasound was used to disperse the particles in the ``feed-suspension''.

\begin{figure}[h]
    \centering
    \includegraphics[width=1.0\textwidth]{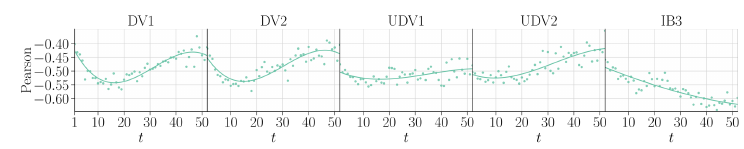}
    \caption{Empirical Pearson correlation coefficients of $d(P)$ and $a(P)$  for all time steps $t\in T$ and  the experiments $\DVa$, $\DVb$, $\UDVa$, $\UDVb$ and $\IBc$.   Additionally, for visualization purposes, fitted polynomials of degree 3 are shown.
    }
    \label{fig:corr_coeff}
\end{figure}

For estimating the bivariate probability densities $f_\abbreviationPP$ and $f_\abbreviationAGG$ we first fitted univariate parametric probability densities for the respective components, see Section~\ref{sec:mod_single}. If the particle descriptors $d(P)$ and $a(P)$ represented by these univariate densities can be considered to be independent random variables with Pearson correlation coefficient equal to zero, then their joint probability density is simply  given by the product of the two marginal densities, see Eq.~\eqref{for.pro.den}. However, the empirical Pearson correlation coefficients of $d(P)$ and $a(P)$ shown in Figure~\ref{fig:corr_coeff} are clearly distinct from zero for all time steps $t\in T$  and all  experiments considered in this paper. Thus, we used the copula approach stated in Section~\ref{sec.two.dim} in order
to model the joint distribution of the particle descriptors  $d(P)$ and  $a(P)$.

\subsection{Initial datasets associated with agglomerates}
\subsubsection{Cardinality of points in $D_\abbreviationEND$}

Recall that in Section~\ref{sec:FittingModelParams},  for estimating the probability density $f_\abbreviationAGG$ associated with agglomerates, we required a sufficiently large union $D_\abbreviationEND = \bigcup_{t=t_{\abbreviationEND}}^{52} D_t$, where $t_{\abbreviationEND}=\min\{t\in T: n_t\le 1.05\, n_{52}\}$. Specifically, we expected the highest fraction of agglomerates at time steps close to the end of an experiment. The results stated in Table~\ref{tab:t_end_time_steps} show that the factor $1.05$ in the definition of  $t_{\abbreviationEND}$ is a reasonable choice. It leads to  cardinalities $|D_\abbreviationEND|$, i.e., numbers of particle descriptor vectors in the union sets $D_\abbreviationEND$, 
which are close to the cardinalities $|D_1|$ of $D_1$, see Figure~\ref{fig:number_of_particles}a, and, in particular, 
 much larger than the number of particles $|D_{52}|$ in $D_{52}$, observed at the last time step $t=52$ of the experiments. 

\begin{table}[h]
	\centering
        \begin{tabular}{P{3.6cm}!{\vrule width 2pt}P{1cm}|P{1cm}|P{1cm}|P{1cm}|P{1cm}} 
            experiment & $\DVa$ & $\DVb$ & $\UDVa$ & $\UDVb$ & $\IBc$\\ \hline
            $t_\mathrm{\abbreviationEND}$ & $46$ & $48$ & $46$ & $47$ & $48$\\ \hline
            $|D_\abbreviationEND|$ & $6128$ & $5719$ & $7110$ & $3654$ & $9739$\\ \hline
            $|D_{52}|$ & $853$ & $1110$ & $1018$ & $608$ & $1904$
        \end{tabular}
    \caption{Cardinalities of $D_\abbreviationEND = \bigcup_{t=t_{\abbreviationEND}}^{52} D_t$ and $D_{52}$ for the experiments $\DVa$, $\DVb$, $\UDVa$, $\UDVb$ and $\IBc$.}
    \label{tab:t_end_time_steps}
\end{table}

\subsubsection{Spatial arrangement of points in $D_\abbreviationAGG$}

\begin{figure}[h]
    \centering
    \includegraphics[width=0.7\textwidth]{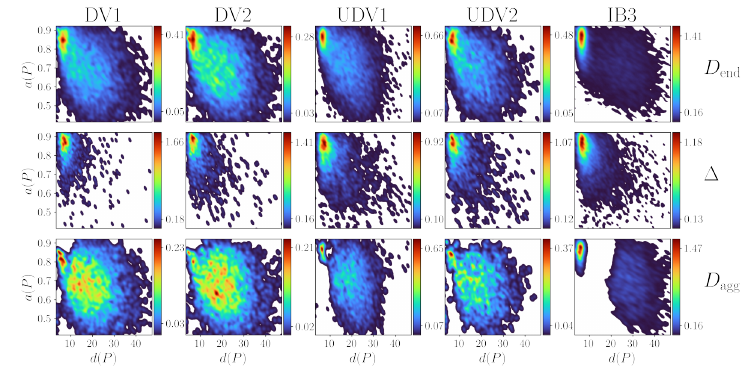}
    \caption{Bivariate probability densities corresponding to the union sets $D_\abbreviationEND$ for the experiments DV1, DV2, UDV1, UDV2 and IB3 (top row), together with densities corresponding to the sets $\deleteSet$ of samples drawn from $f_\abbreviationPP^{(0)}$ (middle row), which are used to get the subsets
    $D_\abbreviationAGG\subset D_\abbreviationEND$ defined in Eq.~(\ref{red.dat.set})  (bottom row).
    }
\label{fig:linear_assignment_results}
\end{figure}

In the next step, after fitting $f_\abbreviationPP^{(0)}$ to the entire set $D_1$, we needed to find a subset of $D_\abbreviationEND$ that ideally consists of all descriptor vectors in $D_\abbreviationEND$ associated with agglomerates. Since no initial fit  $f_\abbreviationAGG^{(0)}$ for the bivariate probability density $f_\abbreviationAGG$ of agglomerates was yet available, we had to determine such a subset $D_\abbreviationAGG\subset D_\abbreviationEND$ by solving the optimization problem stated in Eq.~(\ref{eq:LinearAssignmentProblem}). The bottom row of Figure~\ref{fig:linear_assignment_results} shows  bivariate probability densities corresponding to the subsets $D_\abbreviationAGG$ for the experiments DV1, DV2, UDV1, UDV2 and IB3.

Note that in Figure~\ref{fig:linear_assignment_results} kernel density estimation has been used for  visualizing the datasets $D_\abbreviationEND$,  $\deleteSet$ and  $D_\abbreviationAGG$. It turned out that
in all experiments except for $\DVb$, the largest mode of the  bivariate probability density corresponding to  $D_\abbreviationAGG$ is close to the largest mode of the bivariate probability density corresponding to $\deleteSet$. This is likely due to the continued presence of primary particles in $D_\abbreviationAGG$, which becomes rather obvious in experiment $\IBc$, where $D_\abbreviationAGG$ consists of two clearly distinct clusters of particle descriptor vectors, see Figure~\ref{fig:linear_assignment_results}. Here, the  cluster of descriptor vectors with an area-equivalent diameter of approx. 8 $\si{\micro\meter}$ and an aspect ratio of approx. 0.85 likely corresponds mostly to primary particles, whereas the cluster of descriptor vectors with an area-equivalent diameter of larger than 20 $\si{\micro\meter}$ and an aspect ratio between 0.5 and 0.85 corresponds mostly to agglomerates.

\subsection{Iterative adjustments and resulting fits of  $f_\abbreviationPP^\StarSymbol$ and $f_\abbreviationAGG^\StarSymbol$}

\subsubsection{Spatial arrangement  of points in  the iteratively computed datasets  $D_\abbreviationPP^{(\iteration)}$ and $D_\abbreviationAGG^{(\iteration)}$}

The spatial arrangement of data points in  $D_\abbreviationAGG$, visualized in Figure~\ref{fig:linear_assignment_results}, suggests that
 in most experiments the bivariate probability density $f_\abbreviationAGG^{(0)}$ was likely fitted to a set with a non-negligible fraction of descriptor vectors associated with primary particles. However, the subsequent iterative adjustments discussed in Section~\ref{sec:iterative} were designed to partition $D_1$ as well as $D_\abbreviationEND$ more reliably into sets of descriptor vectors associated with either primary particles or agglomerates, requiring initial estimates $f_\abbreviationPP^{(0)}$ and $f_\abbreviationAGG^{(0)}$ for these adjustments. Throughout the iteration steps $\iteration\in\{1,\ldots,\NumIterations\}$, where different values of $\NumIterations$ have been chosen for each of the experiments  DV1, DV2, UDV1, UDV2 and IB3 according to the rule given in Eq.~(\ref{eq:NumIterations}), see  Table~\ref{tab:NumIterations},
 the kernel density estimates 
 for the bivariate probability density of $D_\abbreviationPP^{(\iteration)}$ exhibited rather small changes.

\begin{table}[h]
	\centering
        \begin{tabular}{P{3.6cm}!{\vrule width 2pt}P{1cm}|P{1cm}|P{1cm}|P{1cm}|P{1cm}} 
            experiment & $\DVa$ & $\DVb$ & $\UDVa$ & $\UDVb$ & $\IBc$\\ \hline
            $\NumIterations$ & $4$ & $7$ & $14$ & $18$ & $30$
        \end{tabular}
    \caption{Number of iterations $\NumIterations$  performed for the experiments $\DVa$, $\DVb$, $\UDVa$, $\UDVb$ and $\IBc$.}
    \label{tab:NumIterations}
\end{table}

\begin{figure}[h]
    \centering
    \includegraphics[width=\textwidth]{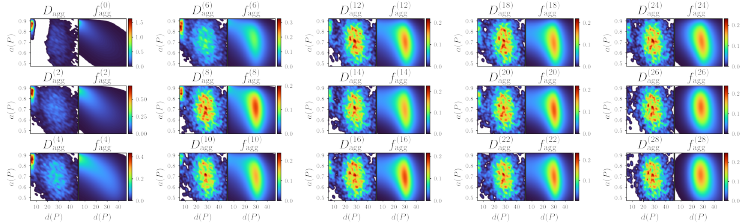}
    \caption{Kernel density estimate of $D_\abbreviationAGG$ and  corresponding bivariate probability density $f_\abbreviationAGG^{(0)}$ for experiment $\IBc$ (top left),  followed by  the result obtained for $D_\abbreviationAGG^{(\iteration)}$ and $f_\abbreviationAGG^{(\iteration)}$ for  all even  iteration steps $\iteration\in\{2,\ldots,28\}$.}
    \label{fig:IB3_f_agg_development_col_combined}
\end{figure}

 On the other hand, the experiments  $\DVa$, $\DVb$ and $\IBc$ showed pronounced changes in the kernel density estimates applied to $D_\abbreviationAGG^{(\iteration)}$. We attribute this behavior to the presumed low fraction of descriptor vectors associated with agglomerates in $D_1$ and the  higher fraction of descriptor vectors associated with primary particles in the initial subset $D_\abbreviationAGG$ of $D_\abbreviationEND$. 
 Thus, there is larger potential for finding subsets of $D_\abbreviationEND$ with increased representativeness of agglomerates and, accordingly, the fits $f_\abbreviationAGG^{(\iteration)}$ also change more. As a result of this, the consecutive adjustments lead to subsets $D_\abbreviationAGG^{(\iteration)}$ that likely contain a reduced fraction of descriptor vectors associated with primary particles as can be seen in Figure~\ref{fig:IB3_f_agg_development_col_combined} for experiment $\IBc$, where with an increasing value of $\iteration$,  the kernel density estimates of the bivariate probability densities corresponding to $D_\abbreviationAGG^{(\iteration)}$ tend to have lower values for descriptor vectors $(d(P), a(P))$ close to the mode at approximately $(8, 0.85)$ that we discussed above. 
 
 Note that there are different scales of the color bars in Figure~\ref{fig:IB3_f_agg_development_col_combined}. Interestingly,  the subset $D_\abbreviationAGG^{(28)}$ in the last even iteration step $\iteration=28$ of experiment $\IBc$ contains hardly any descriptor vectors close to $(8, 0.85)$. Moreover, for each subset $D_\abbreviationAGG^{(\iteration)}$, Figure~\ref{fig:IB3_f_agg_development_col_combined} shows the corresponding  fit  $f_\abbreviationAGG^{(\iteration)}$, where the quality of a fit $f_\abbreviationAGG^{(\iteration)}$ in terms of fitting the corresponding data in $D_\abbreviationAGG^{(\iteration)}$ seems to improve in the course of iterations. For the first iterations, the main obstacle for a good fit is the bimodal distribution of the area-equivalent diameter in $D_\abbreviationAGG^{(\iteration)}$ that is not well fitted by $f_{\abbreviationAGG, d}^{(\iteration)}$. However, during these first iterations, for values of $x=(d(P), a(P))$ close to $(8, 0.85)$, i.e., descriptor vectors likely associated with primary particles, the values of $f_{\abbreviationAGG}^{(\iteration)}(x)$ tend to be smaller than the respective kernel density estimate of $D_\abbreviationAGG^{(\iteration)}$ evaluated at $x$ suggests. This, in turn, leads to a higher estimated probability $\abbreviationPP^{(\iteration +1)}_\BeginOrEnd(x)$ of a descriptor vector $x\in D_\abbreviationEND$ close to $(8, 0.85)$ being associated with a primary particle, see Eqs.~(\ref{eq:probability_iteration_z}) and (\ref{eq:class1_z}). Finally, this leads to the subsequent subset $D_\abbreviationAGG^{(\iteration +1)}$ having a lower density for descriptor vectors close to $(8, 0.85)$ than $D_\abbreviationAGG^{(\iteration)}$.

\subsubsection{Estimates $\widehat{\MixingRatio}_1^{(\iteration)}$ and $\widehat{\MixingRatio}_\abbreviationEND^{(\iteration)}$ with corresponding rank sum  $\rho (\iteration)$ as measure of goodness of fit}\label{sec:estimates}

For the construction of the subsets $D_\abbreviationPP^{(\iteration)}$ and $D_\abbreviationAGG^{(\iteration)}$, explained in Section~\ref{sec:procedure_subseq_adj}, not only the bivariate probability densities $f_\abbreviationPP^{(\iteration)}$ and  $f_\abbreviationAGG^{(\iteration)}$ were used, but also the  estimates $\widehat{\MixingRatio}_1^{(\iteration)}$ and $\widehat{\MixingRatio}_\abbreviationEND^{(\iteration)}$ for $m_1$ and $m_\abbreviationEND$ of the current  iteration step $\iteration\in\{1, \ldots, \NumIterations\}$. Furthermore, the stochasticity introduced by sampling from uniformly distributed random variables on the interval $[0,1]$ adds another layer of complexity to the computation of $D_\abbreviationPP^{(\iteration)}$ and  $D_\abbreviationAGG^{(\iteration)}$.

\begin{figure}[h]
    \centering
    \includegraphics[width=0.85\textwidth]{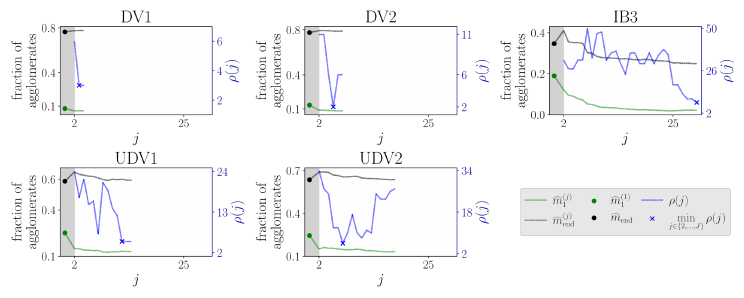}
    \caption{Estimates $\widehat{\MixingRatio}_1^{(1)}$ and $\widehat{\MixingRatio}_\abbreviationEND$ (dots), together with  subsequent estimates $\widehat{\MixingRatio}_1^{(\iteration)}$ (green) and $\widehat{\MixingRatio}_\abbreviationEND^{(\iteration)}$ (black) of the fraction of agglomerates in $D_1$ and $D_\abbreviationEND$, for each iteration step $\iteration\in \{2,\ldots,\NumIterations\}$. The corresponding rank sum $\rho (\iteration)$ is shown in blue, where an additional $\mathrm{y}$-axis on the right-hand side of each plot is given.}
    \label{fig:mixing_ratios_and_rank_per_iteration}
\end{figure}

This complexity makes it challenging to fully grasp the interplay of these factors when constructing $D_\abbreviationPP^{(\iteration)}$ and  $D_\abbreviationAGG^{(\iteration)}$. However, Figure~\ref{fig:mixing_ratios_and_rank_per_iteration} attempts to shed some light on this matter, where for each experiment the estimates $\MixingRatio^{(\iteration)}_1$ and $\MixingRatio^{(\iteration)}_\abbreviationEND$ as well as the rank sum $\rho (\iteration)$ in each iteration step $j\in\{2, \ldots, \NumIterations\}$ is shown. Moreover, the mixing ratios  $\widehat{\MixingRatio}^{(1)}_1$ and $\widehat{\MixingRatio}_\abbreviationEND$  estimated by means of quantiles are included in  gray boxes, whereas the mixing ratios $\MixingRatio^{(\iteration)}_1$ and $\MixingRatio^{(\iteration)}_\abbreviationEND$, which are   obtained by  maximum-likelihood estimation as stated in Eq.~(\ref{eq:MixingRatioML}),   are visualized on the white background.

Recall that the number of iterations $\NumIterations$   is always less than 50, see Eq.~(\ref{eq:NumIterations}). Furthermore, in each experiment the termination condition introduced in Section~\ref{sec:procedure_subseq_adj} is met. This means that at some iteration step  the adjustment of the estimates $\widehat{\MixingRatio}^{(\iteration-1)}_1$ and $\widehat{\MixingRatio}^{(\iteration-1)}_\abbreviationEND$ from the previous iteration becomes marginal. For experiments $\DVa$ and $\DVb$ this happens at early iteration steps $\iteration=4$ and $\iteration=7$, respectively. In all iteration steps of both experiments, the estimates for the mixing ratios hardly change, both in absolute and relative terms. On the other hand, the numbers of adjustments $\NumIterations$ in experiments $\UDVa$ and $\UDVb$ are higher with $\NumIterations=14$ and $\NumIterations=18$, respectively. Moreover, in comparison to experiments $\DVa$ and $\DVb$, the mixing ratios change only slightly more. However, the adjustments of $\widehat{\MixingRatio}_\abbreviationEND$ to $\widehat{\MixingRatio}_\abbreviationEND^{(2)}$ and $\widehat{\MixingRatio}_1^{(1)}$ to $\widehat{\MixingRatio}_1^{(2)}$ are quite considerable. Finally, in experiment $\IBc$, the number of adjustments $\NumIterations=30$ is the highest and the changes of mixing ratios are   most significant in the first iteration steps, whereas the changes become smaller in later iterations. Overall, the large differences between $\widehat{\MixingRatio}_\abbreviationEND$ and $\widehat{\MixingRatio}_\abbreviationEND^{(\iteration^\StarSymbol)}$ or between $\widehat{\MixingRatio}_1^{(1)}$ and $\widehat{\MixingRatio}_1^{(\iteration^\StarSymbol)}$, occurring in some cases,  highlight the necessity of these iterative adjustments.

Since neither the mixing ratios $\widehat{\MixingRatio}_1^{(\iteration)}$ and $\widehat{\MixingRatio}_\abbreviationEND^{(\iteration)}$  nor the rank sum $\rho (\iteration)$ behave monotonously in $\iteration$, it is difficult to evaluate the quality of each of the changes. In experiment $\UDVb$, for example, after iteration $\argmin_{\iteration\in\{2,\ldots,\NumIterations\}}\rho(\iteration)=7$ many iterations follow, in which $\rho (\iteration)$ tends to increase. Although, as we discussed above  in the context of Figure~\ref{fig:IB3_f_agg_development_col_combined}, the subsets $D_\abbreviationAGG^{(\iteration)}$ in experiment $\IBc$ seem to drastically improve over the course of  iterations, the rank sum $\rho(\iteration)$  is clearly not monotonously decreasing. Therefore, we  directly check whether $\iteration^\StarSymbol$, i.e., the argument of the minimum of $\rho(\iteration)$ minus 1, was nonetheless a good choice for determining $f_\abbreviationPP^\StarSymbol(=f_\abbreviationPP^{(\iteration^\StarSymbol)})$ and $f_\abbreviationAGG^\StarSymbol(=f_\abbreviationAGG^{(\iteration^\StarSymbol)})$.

\subsubsection{Final parametric fits $f_\abbreviationPP^\StarSymbol$ and $f_\abbreviationAGG^\StarSymbol$ for primary particles and agglomerates}

\begin{figure}[h]
    \centering
    \includegraphics[width=0.65\textwidth]{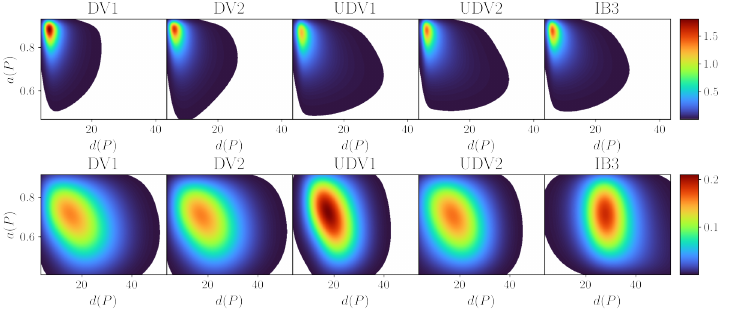}
    \caption{ Visualization of the fitted bivariate probability densities  $f_\abbreviationPP^\StarSymbol$ (top row)  and $f_\abbreviationAGG^\StarSymbol$ (bottom row) for the experiments DV1, DV2, UDV1, UDV2 and IB3.}
    \label{fig:comparison_pp_and_agg}
\end{figure}

In Figure~\ref{fig:comparison_pp_and_agg}, the bivariate probability densities $f_\abbreviationPP^\StarSymbol$ and $f_\abbreviationAGG^\StarSymbol$ are illustrated for each experiment given in Table~\ref{tab:experiments}. As detailed in Section~\ref{sec:consistency_mixture_components}, these densities are subject to certain consistency conditions across the experiments  DV1, DV2, UDV1, UDV2 and IB3, which we can now verify. First, the probability densities $f_\abbreviationPP^\StarSymbol$ across these experiments appear notably similar, confirming our expectation based on the use of the same particle type in each experiment. However, while the modes of all densities $f_\abbreviationPP^\StarSymbol$ are  located around $(8, 0.85)$, the value of $f_\abbreviationPP^\StarSymbol$ evaluated at its mode is significantly higher for experiment $\DVa$. At the same time, the values of $f_\abbreviationPP^\StarSymbol$ for experiment $\DVa$ are smaller for descriptor vectors with a large area-equivalent diameter and a small aspect ratio. Nevertheless, overall the probability densities appear to be very similar as desired. Moreover, the probability densities $f_\abbreviationAGG^\StarSymbol$ for experiments $\DVa$ and $\DVb$ are almost identical, as expected, given that these experiments are repetitions of each other. However, for the other pair of repeated experiments, $\UDVa$ and $\UDVb$, while the modes of their respective fits $f_\abbreviationAGG^\StarSymbol$ are located similarly, the variance of the area-equivalent diameter is smaller and the variance of the aspect ratio is larger for $f_\abbreviationAGG^\StarSymbol$ of $\UDVa$, compared to $f_\abbreviationAGG^\StarSymbol$ of $\UDVb$. Notably, the bivariate density $f_\abbreviationAGG^\StarSymbol$ of $\IBc$ is  most different from the fits of the other experiments, since it exhibits agglomerates that are larger.

For the bivariate probability densities depicted in Figure~\ref{fig:comparison_pp_and_agg}, the parametric families of both, the two marginal densities of area-equivalent diameter and aspect ratio, and the copula, along with the values of their respective fitted parameters, are listed in Table~\ref{tab:parameters_pp_agg_fits}. Except for the probability density of the generalized hyperbolic distribution, which is given in Eq.~(\ref{eq.gen.par}), the probability densities of the other parametric families of univariate distributions appearing in Table~\ref{tab:parameters_pp_agg_fits} are given in Table~\ref{tab:parametric_distribution}.

\begin{table}[h]\small
	\centering
	\begin{tabular} {l | l| l | l}
             \multirow{3}{*}{\parbox{1.8cm}{density/\\ experiment}} & \multirow{3}{*}{\parbox{1.8cm}{particle\\ descriptor}} & \multirow{3}{*}{\parbox{2.5cm}{ marg. distrib./\\ copula type}} & \multirow{3}{*}{fitted parameter values}\\
            & & & \\
            & & & \\
            \specialrule{2pt}{0em}{0em}
            % PP, DV1
            \multirow{3}{*}{$f_\mathrm{pp}^\ast,\mathrm{DV1}$} & area-eq. dia. & gen. hyperbolic & $\alpha=0.9047,\beta=0.8932,\delta=2.7427,\lambda=-1.8455,\mu=5.4743$ \\
            \cline{2-4}
             & aspect ratio & gen. hyperbolic & $\alpha=93.8323,\beta=-84.3239,\delta=0.0373,\lambda=0.0805,\mu=0.9255$ \\
            \cline{2-4}
             &  & Clayton & $\theta=0.4894,r=90$ \\
             \specialrule{2pt}{0em}{0em}
             % PP, DV2
            \multirow{3}{*}{$f_\mathrm{pp}^\ast,\mathrm{DV2}$} & area-eq. dia. & gen. hyperbolic & $\alpha=1.0632,\beta=1.0439,\delta=2.7612,\lambda=-1.6566,\mu=4.9977$ \\
            \cline{2-4}
             & aspect ratio & gen. hyperbolic & $\alpha=71.3184,\beta=-60.9841,\delta=0.0254,\lambda=0.8117,\mu=0.9154$ \\
            \cline{2-4}
             &  & Ali-Mikhail-Haq & $\theta=0.8626, r=90$ \\
             \specialrule{2pt}{0em}{0em}
            % PP, UDV1
            \multirow{3}{*}{$f_\mathrm{pp}^\ast,\mathrm{UDV1}$} & area-eq. dia. & gen. hyperbolic & $\alpha=0.6736,\beta=0.6212,\delta=1.8247,\lambda=-0.3727,\mu=6.0196$ \\
            \cline{2-4}
             & aspect ratio & exp. Weibull & $\alpha=39.0851,\theta=0.2351,\sigma=1.0722,\mu=-0.1818$ \\
            \cline{2-4}
             &  & Ali-Mikhail-Haq & $\theta=0.9154,r=90$ \\
             \specialrule{2pt}{0em}{0em}
            % PP, UDV2
            \multirow{3}{*}{$f_\mathrm{pp}^\ast,\mathrm{UDV2}$} & area-eq. dia. & gen. hyperbolic & $\alpha=0.8722,\beta=0.8147,\delta=1.5869,\lambda=-0.3266,\mu=5.7841$ \\
            \cline{2-4}
             & aspect ratio & exp. Weibull & $\alpha=44.1723,\theta=0.1915,\sigma=0.9934,\mu=-0.0934$ \\
            \cline{2-4}
             &  & Ali-Mikhail-Haq & $\theta=0.9458 ,r=90$ \\
             \specialrule{2pt}{0em}{0em}
             % PP, IB3
            \multirow{3}{*}{$f_\mathrm{pp}^\ast,\mathrm{IB3}$} & area-eq. dia. & gen. hyperbolic & $\alpha=0.8009,\beta=0.7421,\delta=1.7527,\lambda=-0.4319,\mu=5.6903$ \\
            \cline{2-4}
             & aspect ratio & exp. Weibull & $\alpha= 99.3528,\theta=0.2888,\sigma=2.7903,\mu=-1.8971$ \\
            \cline{2-4}
             &  & Ali-Mikhail-Haq & $\theta=0.9249 ,r=90$ \\
             \specialrule{2pt}{0em}{0em}
             % AGG, DV1
            \multirow{3}{*}{$f_\mathrm{agg}^\ast,\mathrm{DV1}$} & area-eq. dia. & Burr Type XII & $c=2.2112,k=30.0675,\sigma=90.1895,\mu=1.9944$ \\
            \cline{2-4}
             & aspect ratio & beta & $\alpha=4.4876,\beta=3.0262,\sigma=0.6849,\mu=0.2693$ \\
            \cline{2-4}
             &  & Ali-Mikhail-Haq & $\theta=-0.8795 ,r=180$ \\
             \specialrule{2pt}{0em}{0em}
             % AGG, DV2
            \multirow{3}{*}{$f_\mathrm{agg}^\ast,\mathrm{DV2}$} & area-eq. dia. & Burr Type XII & $c=2.4661,k=14.7549,\sigma=60.9572,\mu=1.9904$ \\
            \cline{2-4}
             & aspect ratio & beta & $\alpha=4.3865,\beta=2.9258,\sigma=0.6719,\mu=0.2724$ \\
            \cline{2-4}
             &  & Ali-Mikhail-Haq & $\theta=-0.7788 ,r=180$ \\
             \specialrule{2pt}{0em}{0em}
             % AGG, UDV1
            \multirow{3}{*}{$f_\mathrm{agg}^\ast,\mathrm{UDV1}$} & area-eq. dia. & power normal & $\alpha=0.1236,\sigma=2.924, \mu=11.5114$ \\
            \cline{2-4}
             & aspect ratio & beta & $\alpha=3.7349,\beta=2.7293,\sigma=0.6281,\mu=0.3343$ \\
            \cline{2-4}
             &  & Ali-Mikhail-Haq & $\theta= -0.9584,r=180$ \\
             \specialrule{2pt}{0em}{0em}
             % AGG, UDV2
            \multirow{3}{*}{$f_\mathrm{agg}^\ast,\mathrm{UDV2}$} & area-eq. dia. & rice & $\lambda=1.6661,\sigma=8.7099,\mu=1.0592$ \\
            \cline{2-4}
             & aspect ratio & Johnson $\mathrm{S}_\mathrm{b}$ & $\gamma=-0.3208,\delta=1.3273,\sigma=0.6877,\mu=0.3009$ \\
            \cline{2-4}
             &  & Frank & $\theta=-1.1435 ,r=0$ \\
             \specialrule{2pt}{0em}{0em}
             % AGG, IB3
            \multirow{3}{*}{$f_\mathrm{agg}^\ast,\mathrm{IB3}$} & area-eq. dia. & Johnson $\mathrm{S}_\mathrm{U}$ & $\alpha=-0.5473, \beta=1.7638, \sigma=11.6947, \mu=25.2372$ \\
            \cline{2-4}
             & aspect ratio & beta & $\alpha=2.9333,\beta=2.3599,\sigma=0.5718,\mu=0.372$ \\
            \cline{2-4}
             &  & Clayton & $\theta=0.2155 ,r=90$ 
        \end{tabular}
    \caption{Parametric families of marginal distributions and copula types of  $f_\abbreviationPP^\StarSymbol$ and $f_\abbreviationAGG^\StarSymbol$, 
     for the experiments DV1, DV2, UDV1, UDV2 and IB3, together with the fitted parameter values (gen. = generalized, exp. = exponentiated).}
    \label{tab:parameters_pp_agg_fits}
\end{table}

\begin{table}[h!]
	\centering
	\begin{tabular} {p{3.5cm} | p{7cm}| p{1.5cm} | p{4cm}}
             parametric family & probability density & support & parameters\\
            \specialrule{2pt}{0em}{0em}
            exp. Weibull \cite{2020SciPy,Mudholkar1993} & $\sigma^{-1}\alpha\theta(1-\euler^{-(\frac{x-\mu}{\sigma})^\alpha})^{\theta-1}\euler^{-(\frac{x-\mu}{\sigma})^\alpha}(\frac{x-\mu}{\sigma})^{\alpha-1}$ & $[\mu,\infty)$ & $\mu\in\R$ and $\alpha,\theta,\sigma\in (0,\infty)$\\
            \hline
            Burr Type XII \cite{2020SciPy,Tadikamalla1980} & $\sigma^{-1}ck(\frac{x-\mu}{\sigma})^{c-1}(1+(\frac{x-\mu}{\sigma})^{c})^{-(k+1)}$ & $[\mu,\infty)$ & $\mu\in\R$ and $c,k,\sigma\in (0,\infty)$\\
            \hline
            beta \cite{2020SciPy,Johnson1995} & $\frac{(\frac{x-\mu}{\sigma})^{\alpha-1}(1-\frac{x-\mu}{\sigma})^{\beta-1}}{\sigma \mathrm{B}(\alpha,\beta)}$ & $[\mu,\mu +1]$ & $\mu\in\R$ and $\alpha,\beta,\sigma\in (0,\infty)$\\
            \hline
            power normal \cite{2020SciPy, Gupta2008} & $\sigma^{-1}\alpha\phi(\frac{x-\mu}{\sigma})\Phi^{\alpha-1}(-\frac{x-\mu}{\sigma})$ & $\R$ & $\mu\in\R$ and $ \alpha,\sigma\in (0,\infty)$\\
            \hline
            Rice \cite{2020SciPy, Talukdar1991} & $\frac{x-\mu}{\sigma^2}\euler^{-\frac{1}{2}((x-\mu)^2/ \sigma^2+\lambda^2)}\mathrm{B}_0(\frac{x-\mu}{\sigma}\lambda)$ & $(\mu,\infty)$ & $\mu\in\R$ and $ \lambda,\sigma\in (0,\infty)$\\
            \hline
            Johnson $\mathrm{S}_\mathrm{b}$ \cite{2020SciPy,Siekierski1992} & $\frac{\delta\sigma}{(x-\mu)(\sigma-x+\mu)}\phi(\gamma+\delta \log \frac{x-\mu}{\sigma-x +\mu})$ & $(\mu,\mu +1)$ & $\gamma,\mu\in\R$ and $\delta,\sigma\in (0,\infty)$ \\
            \hline
            Johnson $\mathrm{S}_\mathrm{U}$ \cite{2020SciPy,Gunduz2020} & $\frac{\beta}{\sqrt{(x-\mu)^2+\sigma^2}}\phi(\alpha+\beta\sinh^{-1}\frac{x-\mu}{\sigma})$ & $\R$ & $\alpha,\mu\in\R$ and $\beta,\sigma\in (0,\infty)$
        \end{tabular}
    \caption{Parametric families of univariate distributions with corresponding density, support and parameter space. The symbols $\mathrm{B},\mathrm{B}_0,\sinh^{-1}$ denote the beta function \cite{Abramowitz1968}, modified Bessel function of the first kind with index 0 \cite{Abramowitz1968} and the inverse of the hyperbolic sine function, whereas $\phi$ and $\Phi$ denote the probability density and cumulative distribution function of the standard normal distribution.
    }
    \label{tab:parametric_distribution}
\end{table}

\subsection{Model parameters for  the time-dependent
 fraction of agglomerates}\label{sec:ResultsModelParamsMixingRatio}

In Section~\ref{sec:est_frac_agg} we fitted a parametric function $\pmrFunction_{\pmrMinimum,\pmrMaximum,\pmrContBerParam}:T\to[a,b]$, see Eq.~(\ref{eq:pmr}), to the estimated fractions $\widehat{\MixingRatio}_t$ of agglomerates for each time step $t\in T$ by solving the non-linear least squares problem stated in Eq.~(\ref{eq:non-linear-least-squares}).  The small values of $\mathrm{MSE}$ given in
Table~\ref{tab:params_mixing_ratio}
show that the parametric function $\pmrFunction_{\pmrMinimum,\pmrMaximum,\pmrContBerParam}:T\to[a,b]$ introduced in Eq.~(\ref{eq:pmr}) fits  the estimated fractions $\widehat{\MixingRatio}_t$  of agglomerates for all $t\in T$
 quite well.

\begin{table}[h]\small
	\centering
	\begin{tabular} {l | l| l| l| l}
             experiment & $\pmrMinimum^\StarSymbol$ & $\pmrMaximum^\StarSymbol$ & $\pmrContBerParam^\StarSymbol$ & $\mathrm{MSE}$ \\
            \specialrule{2pt}{0em}{0em}
            $\DVa$ & 0.0129 & 0.8074 & 0.0810 & $5.217\cdot 10^{-4}$ \\
            \hline
            $\DVb$ & 0.0163 & 0.8217 & 0.1105 & $7.211\cdot 10^{-4}$ \\
            \hline
            $\UDVa$ & 0.1027 & 0.5978 & 0.0146 & $3.029\cdot 10^{-4}$ \\
            \hline
            $\UDVb$ & 0.1243 & 0.6749 & 0.1190 & $3.820\cdot 10^{-4}$ \\
            \hline
            $\IBc$ & 0.0078 & 0.2543 & 0.3013 & $8.091\cdot 10^{-5}$
        \end{tabular}
    \caption{Optimal parameter values $\pmrMinimum^\StarSymbol,\pmrMaximum^\StarSymbol$ and $\pmrContBerParam^\StarSymbol$
    of the function $\pmrFunction_{\pmrMinimum,\pmrMaximum,\pmrContBerParam}:T\to[a,b]$ introduced in Eq.~(\ref{eq:pmr}), together with the correspnoding $\mathrm{MSE}$, for the experiments DV1, DV2, UDV1, UDV2 and IB3.
    }
    \label{tab:params_mixing_ratio}
\end{table}

Recall that the optimal parameter values $\pmrMinimum^\StarSymbol,\pmrMaximum^\StarSymbol,\pmrContBerParam^\StarSymbol$ provided in 
Table~\ref{tab:params_mixing_ratio}
are closely related with the dynamics of the agglomeration experiments. For example, $\pmrMinimum^\StarSymbol$ and $\pmrMaximum^\StarSymbol$ represent the estimated fractions $\widehat{\MixingRatio}_1$ and $\widehat{\MixingRatio}_{52}$ of agglomerates at the beginning and end of the experiment, i.e.,  $\widehat{\MixingRatio}_1\approx\pmrFunction_{\pmrMinimum^\StarSymbol,\pmrMaximum^\StarSymbol,\pmrContBerParam^\StarSymbol}(1)=\pmrMinimum^\StarSymbol$ and $\widehat{\MixingRatio}_{52}\approx\pmrFunction_{\pmrMinimum^\StarSymbol,\pmrMaximum^\StarSymbol,\pmrContBerParam^\StarSymbol}(52)=\pmrMaximum^\StarSymbol$. Moreover, in Section~\ref{sec:eval_a_b_lambda} we discussed some consistencies between $\pmrMinimum^\StarSymbol,\pmrMaximum^\StarSymbol$ and $\pmrContBerParam^\StarSymbol$ which we verify in the following. Namely, the fraction of agglomerates at the beginning of the experiments,  represented by $\pmrMinimum^\StarSymbol$, should be similar for the experiments $\UDVa$ and $\UDVb$, as well as for  $\DVa, \DVb$ and $ \IBc$. Furthermore, the value of $\pmrMinimum^\StarSymbol$ should be larger for $\UDVa$ and $\UDVb$ than for the remaining experiments. Remarkably, the mean value of $\pmrMinimum^\StarSymbol$ for experiments $\DVa, \DVb$ and $ \IBc$ is only slightly above 0 (with a very  small mean agglomerate fraction  of 1.23\%), whereas the mean value of $\pmrMinimum^\StarSymbol$ for experiments $\UDVa$ and $\UDVb$, in which no ultrasound was used, is significantly higher (with a mean agglomerate fraction of 11.35\%). The use or non-use of ultrasound is the only difference between the pairs of repeated experiments $\DVa,\DVb$ and $\UDVa,\UDVb$, which not only leads to  lower values of $\pmrMinimum^\StarSymbol$ in $\DVa$ and $\DVb$, but also to  higher values of $\pmrMaximum^\StarSymbol$ than in the experiments $\UDVa$ and $\UDVb$.

As stated above, we  checked the similarity of the values of $\pmrMinimum^\StarSymbol, \pmrMaximum^\StarSymbol$ and $\pmrContBerParam^\StarSymbol$ for the repeated experiments $\DVa$ and $\DVb$, as well as for $\UDVa$ and $\UDVb$. For  $\DVa$ and $\DVb$ these values are quite similar to each other,
whereas  the values
of $\pmrMinimum^\StarSymbol, \pmrMaximum^\StarSymbol$ and $\pmrContBerParam^\StarSymbol$  behave 
differently for  $\UDVa$ and $\UDVb$,  see 
Table~\ref{tab:params_mixing_ratio}. While the difference between the two values of  $\pmrMinimum^\StarSymbol$ is also small for  $\UDVa$ and $\UDVb$, the differences for $\pmrMaximum^\StarSymbol$ and $\pmrContBerParam^\StarSymbol$ are larger. 
Recall that in  Figure~\ref{fig:cdf_continuous_bernoulli} we discussed the influence of the parameter $\pmrContBerParam$ on the shape of the function $\pmrFunction_{\pmrMinimum,\pmrMaximum,\pmrContBerParam}:T\to[a,b]$. In particular,  the value of $\pmrContBerParam^\StarSymbol=0.1190\approx 0.1$ obtained for $\UDVb$ 
indicates a slower increase of the fraction of agglomerates at the beginning and a stronger increase towards the end of the agglomeration experiment compared to the value of $\pmrContBerParam^\StarSymbol=0.0146\approx 0.01$ obtained for $\UDVa$.
Finally, note that the largest value of $\pmrContBerParam^\StarSymbol=0.3013$ has been obtained for experiment $\IBc$, which differs clearly from the values of $\pmrContBerParam^\StarSymbol$ for the remaining experiments. For $\pmrContBerParam^\StarSymbol=0.3013$,
there is still a non-negligible gradient at $t=52$, in contrast to the situation observed for the remaining experiments at a higher energy dissipation rate, see Figure~\ref{fig:cdf_continuous_bernoulli}. This suggests the possibility that the increase in the fraction of agglomerates has not yet reached saturation at the end of  experiment $\IBc$, which is a consequence of the lower energy dissipation rate for IB3, and the therewith associated lower probability of collision between the particles.

\subsection{Distribution of the time-dependent content in the stirred tank}\label{sec.con.tan}

\begin{figure}[h!]
     \centering
     \includegraphics[width=0.6\textwidth]{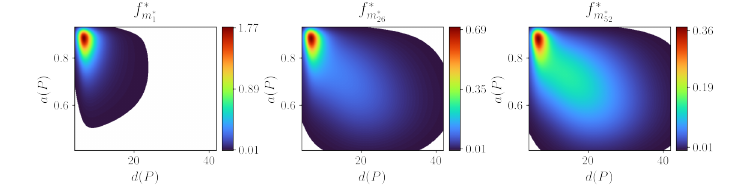}
     \caption{Fitted bivariate probability densities $f^\StarSymbol_{\MixingRatio_{1}^\StarSymbol}$ (left), $f^\StarSymbol_{\MixingRatio_{26}^\StarSymbol}$ (middle) and $f^\StarSymbol_{\MixingRatio_{52}^\StarSymbol}$ (right) of experiment $\DVa$.}
     \label{fig:fits_t_1_26_52}
 \end{figure}

 For each time  step $t\in T$,
we modeled the bivariate probability density $f^\StarSymbol_{\MixingRatio_t^\StarSymbol}$ of particle descriptor vectors in $D_t$ as a mixture of 
 $f_\abbreviationPP^\StarSymbol$ and $f_\abbreviationAGG^\StarSymbol$, 
 see   Eq.~(\ref{eq:final_fit}) in Section~\ref{sec:est_frac_agg}.
Figure~\ref{fig:fits_t_1_26_52} shows the fitted density $f^\StarSymbol_{\MixingRatio_t^\StarSymbol}$ of experiment $\DVa$ at time steps $t\in\{1,26, 52\}$ or in other words, approximately 0.5, 23.5 and 47 minutes after the ``feed-suspension'' is added to the stirred tank.

\begin{figure}[h]
     \centering
     \includegraphics[width=\textwidth]{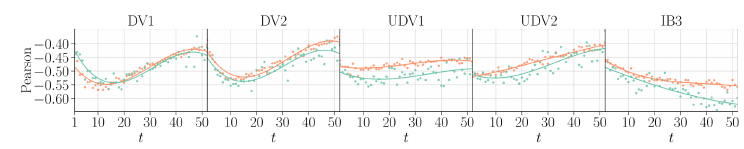}
     \caption{Empirical Pearson correlation coefficients of the particle descriptors $d(P)$ and $a(P)$ in $D_t$ (green dots) and in samples of 10,000 realizations drawn from $f^\StarSymbol_{\MixingRatio_t^\StarSymbol}$ (orange dots),   for each $t\in T$. Additionally, like in Figure~\ref{fig:corr_coeff},
    fitted polynomials of degree 3 are shown in the respective colors.}
     \label{fig:pearson}
 \end{figure}

So far, we evaluated the goodness-of-fit of  the components  $f_\abbreviationPP^\StarSymbol, f_\abbreviationAGG^\StarSymbol$ and $\MixingRatio^\StarSymbol_{t^\StarSymbol} (=\pmrFunction_{\pmrMinimum^\StarSymbol,\pmrMaximum^\StarSymbol,\pmrContBerParam^\StarSymbol}(t))$  of the mixture  $f^\StarSymbol_{\MixingRatio_t^\StarSymbol}$. However, in general,  we can not conclude from this, whether  $f^\StarSymbol_{m_t^\StarSymbol}$ itself fits sufficiently well  to its corresponding dataset $D_t$. Therefore, as explained in Section~\ref{sec.fiv.thr}, we   generate (artificial)  two-dimensional vector data sampled from $f^\StarSymbol_{\MixingRatio_t^\StarSymbol}$ and compare various characteristics of them with the corresponding characteristics computed for the dataset $D_t$. First,  we examine the empirical correlation coefficient between  area-equivalent diameter and  aspect ratio. For experiments $\DVa$, $\DVb$, and $\UDVb$, the values of this characteristic computed from measured and simulated data, respectively, are quite close to each other, while the differences for experiments $\UDVa$ and  $\IBc$  are somewhat larger, see Figure~\ref{fig:pearson}.

\begin{figure}[h]
     \centering
     \includegraphics[width=0.9\textwidth]{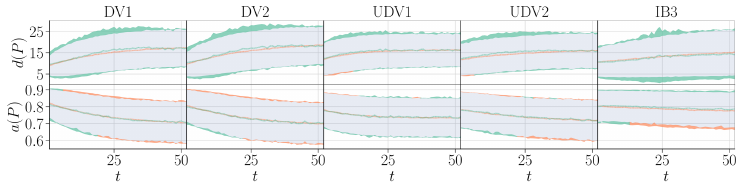}
     \caption{Time-dependent statistics of area-equivalent diameter (top row) and aspect-ratio (bottom row). Specifically, the dashed lines represent the means of the respective descriptor in $D_t$ (green) and in the 10,000 realizations drawn from $f^\StarSymbol_{\MixingRatio_t^\StarSymbol}$ (orange), for each time step $t\in T$. Corresponding standard deviation bands are plotted in the respective colors, where intersections of green and orange bands are visualized in purple.}
     \label{fig:mean_with_sd}
 \end{figure}

In Figure~\ref{fig:mean_with_sd},    mean value functions along with corresponding standard deviation bands are visualized for  area-equivalent diameter and  aspect ratio across all time steps $t\in T$ for the experiments DV1, DV2, UDV1, UDV2 and IB3, where
the purple areas highlight the intersections of the standard deviation bands computed for values of the respective descriptor in $D_t$ (green) and in the simulated dataset drawn from $f^\StarSymbol_{m_t^\StarSymbol}$ (orange). Remarkably, the curves representing the mean values are nearly identical across all instances. Moreover, the standard deviation bands for the aspect ratio computed for the measured datasets $D_t$ and the simulated  data 
drawn from $f^\StarSymbol_{m_t^\StarSymbol}$, respectively, coincide nicely, whereas the standard deviation bands  for the area-equivalent diameter  differ slightly from each other in some cases. However, overall Figure~\ref{fig:mean_with_sd} shows a very good match between the mean value functions and standard deviation bands computed for measured and simulated values of particle descriptors, respectively.
 
\begin{figure}[h]
     \centering
     \includegraphics[width=0.9\textwidth]{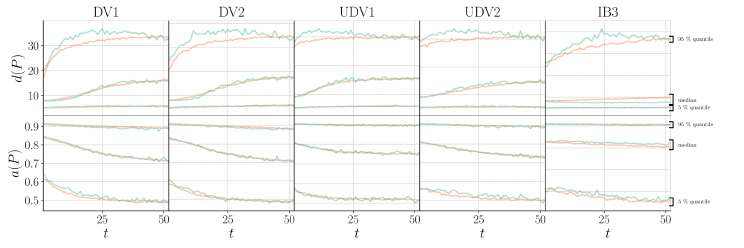}
     \caption{Median, as well as 5 and 95 percent quantiles of
     area-equivalent diameter (top)
     and aspect ratio (bottom)
     computed for $D_t$ (green) and  simulated data drawn from $f^\StarSymbol_{\MixingRatio_t^\StarSymbol}$ (orange), respectively, for each time step $t\in T$.}
     \label{fig:quantiles}
 \end{figure}

 Finally,  we investigate  the 5 percent, 50 percent (median), and 95 percent quantiles of the values of area-equivalent diameter  and aspect-ratio computed from  $D_t$ and simulated data, respectively, for each time step $t\in T$. The obtained results are shown in Figure~\ref{fig:quantiles}. Here, one can see that the  quantiles of  aspect ratio computed from from measured and simulated data are almost identical for each time step $t\in T$ and for each experiment. In addition,  5 percent quantile and  median of  area-equivalent diameter coincide nicely for each time step $t\in T$ and for each experiment except for $\IBc$, where for increasing $t$ the discrepancy between the medians of area-equivalent diameter computed from $D_t$ and simulated data, respectively, tends to monotonously increase to approximately 3.3 \si{\micro\meter}.  The 95 percent quantiles of  area-equivalent diameter computed from measured data 
 in $D_t$ are slightly larger than those computed from simulated data, especially  for time steps $t\in T$ in the middle of the experiments, see
Figure~\ref{fig:quantiles}.

\section{Discussion}\label{sec:discussion}

The present paper introduces a probabilistic modeling approach for hydrophobic agglomeration processes, where the bivariate probability densities $f_\abbreviationPP^\StarSymbol$ and $f_\abbreviationAGG^\StarSymbol$ of the descriptor vectors (of area-equivalent diameter and aspect ratio) of primary particles and agglomerates, respectively, do not depend on the particular time step $t\in T$. 
For  primary particles, this can be justified by the assumption  that primary particles, i.e., single solid particles, do not change their sizes and shapes throughout the agglomeration experiments. On the other hand, the assumption that the joint  density $f_\abbreviationAGG^\StarSymbol$ of area-equivalent diameter and aspect ratio of agglomerates does not change over time
 requires the presence of certain specific circumstances as, for example, the breaking of large agglomerates due to shear forces. This prevents agglomerates from growing unboundedly large, which in turn is necessary for an equilibrium in the size distribution of agglomerates.

The results obtained in this paper show that the fitted mixtures
$f^\StarSymbol_{m_t^\StarSymbol}=(1-\MixingRatio_{t}^\StarSymbol)f_{\abbreviationPP}^\StarSymbol+\MixingRatio_{t}^\StarSymbol f_{\abbreviationAGG}^\StarSymbol$ 
of the densities  $f_\abbreviationPP^\StarSymbol$ and $f_\abbreviationAGG^\StarSymbol$,
with appropriately chosen  (time-dependent) mixing ratios 
$\MixingRatio_{t}^\StarSymbol$,
manage to accurately describe the underlying data in $D_t$  at each of the 52 time steps $t\in T$. 
Specifically, for the  statistics considered in Section~\ref{sec.con.tan}, i.e., for correlation coefficients, quantiles, mean values and standard deviations, only small discrepancies were observed between the respective quantities computed for data from  $D_t$ and  simulated data drawn from  $f^\StarSymbol_{\MixingRatio_t^\StarSymbol}$, for each time step $t\in T$, see Figures~\ref{fig:pearson}, \ref{fig:mean_with_sd} and \ref{fig:quantiles}. Besides this, our  modeling approach  satisfies some necessary consistencies across the experiments, e.g., the fitted probability densities $f_\abbreviationPP^\StarSymbol$ being similar to each other, which is caused  by the fact that all  experiments  use the same type of primary particles, see Figure~\ref{fig:comparison_pp_and_agg}, irrespective of the initial state of particle dispersion.

It should be emphasized that for the following two reasons the model is most likely not overfitted to measured data: on the one hand, the  probability densities $f_\abbreviationPP^\StarSymbol$ and $f_\abbreviationAGG^\StarSymbol$
are parametric and were only fitted by means of data from the first and a few time steps at the end of the agglomeration experiments.
 Consequently, they were not explicitly fitted for all time steps. Nevertheless, the evaluation results show good agreement across all time steps, indicating that the fits capture the distribution underlying the data quite well. A similar validation approach is also commonly used in neural networks to investigate overfitting, as these models are often trained iteratively and---due to their even larger number of parameters---tend to exhibit overfitting more strongly. On the other hand, the time-dependent mixing ratios  
$\MixingRatio_{t}^\StarSymbol$  are modeled by a parametric function 
$\pmrFunction_{\pmrMinimum,\pmrMaximum,\pmrContBerParam}:T\to[a,b]$
with merely three parameters which are  closely related with the dynamics of the agglomeration experiments. Specifically, the parameters $\pmrMinimum$ and $\pmrMaximum$ represent the fractions of agglomerates at the beginning and  the end of the experiment, respectively. The third parameter $\pmrContBerParam$ indicates whether the monotonous increase from $\pmrMinimum$ to $\pmrMaximum$ is linear over the course of the experiments, or whether the increase of the fraction of agglomerates is stronger at the beginning and weaker at the end of the experiment, or vice versa. 
Beyond their role in modeling the agglomeration process, the parameters  $a,b,\lambda$ can be utilized to guide process optimization. For instance,
the low-parametric description of the agglomeration behavior by $a,b,\lambda$
facilitates correlation with relevant process parameters, such as temperature, mixing speed or reactant concentrations, for example, by means of regression. This enables a systematic identification of dependencies between process parameters and the model parameters $a,b,\lambda$. Once these quantitative relationships are established, they can be leveraged for model-based process control and optimization, see, e.g., \cite{furat2020stochastic}.

Recall that some agglomeration experiments considered in this paper dispersed the ``feed-suspension'' by ethanol and ultrasound, namely $\DVa,\DVb$ and $\IBc$, whereas in experiments $\UDVa$ and $\UDVb$ no ultrasound was used. This leads to the assumption that the fitted fraction of agglomerates $\pmrMinimum^\StarSymbol$ at time step $t=1$, i.e., 30-40 seconds after the ``feed-suspension'' is added to the tank, should be higher for $\UDVa$ and $\UDVb$ than for 
$\DVa$, $\DVb$ and $\IBc$.
This assumption is confirmed by the fact that the value obtained for $\pmrMinimum^\StarSymbol$ is approximately equal to 0.01 for experiments, where ultrasound was applied, and approximately equal to 0.11 for experiments, where no ultrasound was applied, see Table~\ref{tab:params_mixing_ratio}. Note that the smallest value for $\pmrMinimum^\StarSymbol$ is 0.0078 and has been obtained for experiment IB3. 

Comparing the results obtained for the pairs of repeated experiments $\DVa$/$\DVb$ and $\UDVa$/$\UDVb$, it is  expected that approximately the same fraction of agglomerates $\pmrMaximum^\StarSymbol$ is present at the ends of each pair of experiments at time step $t=52$, as these experiments use the same energy dissipation rate associated with the Rushton turbine. One could also expect that the value of $\pmrMaximum^\StarSymbol$ could be slightly higher for $\UDVa$/$\UDVb$, since $\pmrMinimum^\StarSymbol$ is larger for these experiments as  mentioned above. Surprisingly, it is the other way around: the mean value of $\pmrMaximum^\StarSymbol$ is smaller for $\UDVa$/$\UDVb$, being equal to   0.64, than for $\DVa$/$\DVb$, where the mean value    of $\pmrMaximum^\StarSymbol$ is  equal to  0.81, see Table~\ref{tab:params_mixing_ratio}. Since the experiments $\DVa$/$\DVb$ and $\UDVa$/$\UDVb$ do not differ from each other except for the use or non-use of ultrasound,  and because their fitted probability densities $f_\abbreviationPP^\StarSymbol$ and $f_\abbreviationAGG^\StarSymbol$ are similar, one can argue that a more dispersed ``feed-suspension'' leads to a higher fraction of agglomerates at the end of the experiment.

This is likely a consequence of more agglomerates, i.e., large particles, being present at the beginning of experiments UDV1 and UDV2. Recall that large particles are more likely to immobilize at the gas-liquid interface, see Section~\ref{sec:segmentation}. As a result, a higher number of primary particles, mostly part of an agglomerate, immobilize over the course of an experiment, where no ultrasound was used. Given that the same quantity of primary particles was used in each of the considered experiments in Table~\ref{tab:experiments}, this indicates a lower number of primary particles (including those in agglomerates) remaining under turbulence, i.e., not being immobilized, at the end of the experiments UDV1 and UDV2 compared to DV1 and DV2. Recall that agglomeration saturates when a certain minimum number/volume-based fraction of particles is reached, see Section~\ref{sec:desc_agglomeration_process}. This fraction is consistent across the experiments DV1, DV2, UDV1 and UDV2 due to the uniform energy dissipation rate of 0.3 and the same volume of the suspension in the stirred tank, see Table~\ref{tab:experiments}. Since the probability densities $f_\abbreviationPP^\StarSymbol$ and $f_\abbreviationAGG^\StarSymbol$ of size and shape descriptors for primary particles and agglomerates, respectively, are similar, the expected number of primary particles in an agglomerate is also similar across the experiments DV1, DV2, UDV1 and UDV2. Consequently, the higher number of primary particles (including those in agglomerates) remaining under turbulence at the end of the experiments DV1 and DV2 in comparison to the experiments UDV1 and UDV2, necessitates a higher fraction $\pmrMaximum^\StarSymbol$ of agglomerates at the respective end. This is required to reach the minimum number/volume-based fraction of particles for which agglomeration saturates.

 Furthermore, one can compare the results obtained for experiment $\IBc$ with those of  $\DVa$ and $\DVb$, which differ in their energy dissipation rates that are associated with the two different stirrer types. Specifically, $\IBc$ used an IB stirrer, whereas the experiments $\DVa$ and $\DVb$ both used a Rushton-turbine. As already mentioned above, the value of $\pmrMinimum^\StarSymbol$ is similar in each case due to the use of ultrasound. However, as we pointed out earlier $\pmrMinimum^\StarSymbol$ is slightly smaller for IB3 than for DV1 and DV2, as expected in Section~\ref{sec:eval_a_b_lambda}, confirming that the higher energy dissipation rate in DV1 and DV2 results in a higher fraction of agglomerates within just 30-40 seconds after the ``feed-suspension'' is added to the tank. Moreover, for experiment $\IBc$ the fitted value of $\pmrMaximum^\StarSymbol$ is equal to  0.25 and thus significantly smaller than the values of $\pmrMaximum^\StarSymbol$ obtained  for experiments $\DVa$ and $\DVb$. However, the fitted value of $\pmrContBerParam^\StarSymbol$ for  $\IBc$ is equal to 0.3, which results in a  function $\pmrFunction_{\pmrMinimum^\StarSymbol,\pmrMaximum^\StarSymbol,\pmrContBerParam^\StarSymbol}$ of mixing ratios that does not seem to be saturating towards a value close to $\pmrMaximum^\StarSymbol$, but a value that is larger than $\pmrMaximum^\StarSymbol$ by a non-negligible amount of (non-available) time steps $t>52$ as mentioned at the end of Section~\ref{sec:ResultsModelParamsMixingRatio}. This suggests extrapolating the mixing ratios $\widehat{\MixingRatio}_t$ for $t>52$ by means of those fitted from Eq.~(\ref{eq:MixingRatioMaxLik}), or running an experiment under the same conditions for a longer time horizon, which could be realized  in  future research. By this, one could check if higher fractions of agglomerates are achievable when using an IB stirrer. For the remaining experiments $\DVa,\DVb,\UDVa$ and $\UDVb$ the fitted values of $\pmrContBerParam^\StarSymbol$ are smaller than 0.12, which in each case indicates a strong increase of the fraction of agglomerates at early time steps and a weak increase at later time steps, with saturation towards the end of the respective experiment.

Regarding the optimization of agglomeration processes, like those investigated in the present paper, not only the maximization of the  fraction  $\pmrMaximum^\StarSymbol$ of agglomerates is of interest, but also ways to design the process such that the resulting agglomerates possess desired size and shape characteristics. Figure~\ref{fig:comparison_pp_and_agg} shows that the bivariate density $f_\abbreviationAGG^\StarSymbol$ of  area-equivalent diameter and  aspect ratio of agglomerates depends heavily on the energy dissipation rate generated by different stirrer types, but much less on the use or non-use of ultrasound. Specifically, flows induced by a Rushton-turbine or an IB stirrer lead to different bivariate densities $f_\abbreviationAGG^\StarSymbol$  of  agglomerates, where two main distinctions can be identified. First, the mode of the area-equivalent diameter is at about 28 \si{\micro\meter} for agglomerates resulting from the usage of an IB stirrer, whereas the modes of the area-equivalent diameter for agglomerates resulting from experiments employing a Rushton-turbine range from 16.5 to 18.5 \si{\micro\meter}. Furthermore, the (negative) correlation of area-equivalent diameter and  aspect ratio of agglomerates resulting from the usage of an IB stirrer is slightly  stronger than for agglomerates resulting from using a Rushton-turbine, see Figure~\ref{fig:corr_coeff}.
However, the marginal distribution of  aspect ratio shows minimal sensitivity to the choice of stirrer type and the associated energy dissipation rate, see the bottom row of Figure~\ref{fig:comparison_pp_and_agg}. On the other hand, the use of an IB stirrer, instead of a Rushton-turbine, results in larger agglomerates, but likely a lower fraction of agglomerates at the end of the experiment.

As indicated in the beginning of this section, our modeling approach can also be applied to other agglomeration processes, where particles can be partitioned into primary particles and agglomerates. However, note that for the experiments considered in the present paper we assumed that no agglomerates are present in the set $D_1$ of descriptor vectors obtained for the first time step $t=1$. Therefore, it is important that the fraction of agglomerates at the beginning of an experiment that is intended to be modeled is not too large, i.e., particles are well dispersed initially. Table~\ref{tab:params_mixing_ratio} shows that the fractions of agglomerates $\pmrMinimum^\StarSymbol$ at the beginning of  experiments  DV1, DV2, UDV1, UDV2 and IB3 are slightly larger than zero, which is contrary to our initial assumption that no agglomerates are present at time step $t=1$. 
Thus, the small fractions  $\pmrMinimum^\StarSymbol$ of agglomerates encompassed within the dataset $D_1$ observed at time step $t=1$ are erroneously considered for fitting $f_\abbreviationPP^{(0)}$. Since we used $f_\abbreviationPP^{(0)}$ to obtain the dataset $D_\abbreviationAGG$ given in
Eq.~(\ref{red.dat.set}), our assumption made regarding time step $t=1$ also affects the fit of $f_\abbreviationAGG^{(0)}$. However, as we have shown, e.g., in Figure~\ref{fig:IB3_f_agg_development_col_combined}, the iterative adjustments explained in Section~\ref{sec:procedure_subseq_adj} can make up for a non-ideal initial fit $f_\abbreviationPP^{(0)}$ created under the slightly perturbated circumstances mentioned above. Specifically, Figure~\ref{fig:IB3_f_agg_development_col_combined} indicates that in the course of iterations subsets of $D_\abbreviationEND$ are found that are more representative for particle descriptor vectors of agglomerates than $D_\abbreviationAGG$. This leads to better fits of the bivariate probability density $f_\abbreviationAGG$ of agglomerates which in turn leads to better fits of the corresponding density 
$f_\abbreviationPP$
of primary particles. Nevertheless, $f_\abbreviationPP^{(0)}$ should be somewhat close to the actual distribution of primary particles, i.e., the fraction of agglomerates at the beginning of an experiment must not be too large.

A key question is whether the methodology presented in this paper can be adapted to more complex particle systems beyond. A relevant extension could be to move from a monodisperse to a polydisperse particle feed, where the descriptors of primary particles follow multimodal distributions. In this case, the parametric model for the primary particles must be adjusted. Instead of a unimodal bivariate probability distribution, a mixture of multiple unimodal bivariate distributions may be required. If such a model can be established, all other steps of the algorithm, including the iterative procedure for distinguishing between primary particles and agglomerates, remain applicable without modification. This suggests that the presented methodology is adaptable to polydisperse particle feeds and potentially to other complex particle systems.

Building up on the methodology developed in this paper, future research may extend the current set of process parameters, namely the energy dissipation rate and the use or non-use of ultrasound, for investigating how various combinations of process parameters impact the  effectiveness of hydrophobic agglomeration processes. This would entail establishing  quantitative relationships that connect process parameters with characteristics of the resulting agglomerates, such as the mean value  or variance of size and shape descriptors as well as the finally achieved fraction of agglomerates. By means of these relationships, the following inverse problem can be studied: for desired specifications of agglomerate characteristics the values of process parameters which are most likely to result in agglomerates exhibiting these predefined characteristics need to be determined. In this way, one can identify promising process parameters without having to carry out a large number of real laboratory experiments.

\section{Conclusion}\label{sec.con.clu}

This study has introduced a suitable probabilistic modeling approach for hydrophobic agglomeration processes, where size and shape descriptors of primary particles and agglomerates follow fixed densities $f_{\abbreviationPP}^\StarSymbol$ and $f_{\abbreviationAGG}^\StarSymbol$, which do not depend on the particular stage of the agglomeration process. The stability of these densities throughout the agglomeration process confirms certain assumptions about process behavior, such as the breaking of large agglomerates due to shear forces, which prevents agglomerates from growing unboundedly large and thus contributes to an equilibrium in the size distribution of particles.

The obtained results demonstrate that the fitted mixtures 
$f^\StarSymbol_{m_t^\StarSymbol}=(1-\MixingRatio_{t}^\StarSymbol)f_{\abbreviationPP}^\StarSymbol+\MixingRatio_{t}^\StarSymbol f_{\abbreviationAGG}^\StarSymbol$  of these densities, with suitably chosen time-dependent mixing ratios $\MixingRatio_{t}^\StarSymbol$, accurately capture the underlying data on  area-equivalent diameter and aspect ratio of particles at each time step. While minor discrepancies were observed for certain statistics, overall, our model performed well in capturing the dynamics of agglomeration processes across various kinds of experiments.

Moreover, the discussion conducted in the previous section highlighted insights into the impact of process parameters, such as the use or non-use of ultrasound and the energy dissipation rate, on the agglomeration process. Specifically, experiments employing a more dispersed ``feed-suspension'' resulted in  higher fractions of agglomerates at the end of the experiments. Furthermore, the energy dissipation rate associated with a specific stirrer type significantly affected the resulting agglomerate characteristics. Flows induced by an IB stirrer led to larger agglomerates with a slightly stronger negative correlation of area-equivalent diameter and  aspect ratio compared to those obtained for a Rushton-turbine. On the other hand, the experiment which used an IB stirrer led to a lower fraction of agglomerates at the end of the experiment, confirming distinct agglomeration dynamics induced by the stirrer type with a corresponding energy dissipation rate.

In view of these findings, future research could focus on extending the set of process parameters and establishing quantitative relationships between process parameters and characteristics of resulting agglomerates. By addressing the inverse problem of agglomeration process optimization, promising specifications of process parameters for achieving desired agglomerate characteristics can be identified without the need for extensive laboratory experiments.
Thus, the present paper not only deepens our understanding of agglomeration dynamics, but also paves the way for practical advancements in industrial agglomeration processes, ultimately leading to cleaner, higher-quality products and more efficient suspension separation methods based on tailored agglomerates.

\section*{Acknowledgement}
The authors gratefully acknowledge funding by the German Research Foundation (DFG) 
within SPP 2364
under grants PE~1160/37-1 and SCHM 997/47-1 and for 
supporting the Collaborative Research Center CRC 920 (Project~ID 
169148856 – Subproject B04).

\bibliography{Bibliography}{}

\begin{thebibliography}{10}
\expandafter\ifx\csname url\endcsname\relax
  \def\url#1{\texttt{#1}}\fi
\expandafter\ifx\csname urlprefix\endcsname\relax\def\urlprefix{URL }\fi
\expandafter\ifx\csname href\endcsname\relax
  \def\href#1#2{#2} \def\path#1{#1}\fi

\bibitem{Taniguchi1996}
S.~Taniguchi, A.~Kikuchi, T.~Ise, N.~Shoji, Model experiment on the coagulation
  of inclusion particles in liquid steel, ISIJ International 36 (1996)
  117--120.

\bibitem{Knupfer2017}
P.~Kn{\"u}pfer, L.~Ditscherlein, U.~A. Peuker, Nanobubble enhanced
  agglomeration of hydrophobic powders, Colloids and Surfaces A:
  Physicochemical and Engineering Aspects 530 (2017) 117--123.

\bibitem{Heuzeroth2015}
F.~Heuzeroth, J.~Fritzsche, U.~A. Peuker, Wetting and its influence on the
  filtration ability of ceramic foam filters, Particuology 18 (2015) 50--57.

\bibitem{Nicklas2023}
J.~Nicklas, U.~A. Peuker, Agglomeration of fine hydrophobic particles: {1D} and
  {2D} characterization by dynamic image analysis of in-line probe data, Powder
  Technology 426 (2023) 118685.

\bibitem{Gruy2005}
F.~Gruy, M.~Cournil, P.~Cugniet, Influence of nonwetting on the aggregation
  dynamics of micronic solid particles in a turbulent medium, Journal of
  Colloid and Interface Science 284 (2005) 548--559.

\bibitem{Van2017}
K.~van Netten, D.~J. Borrow, K.~P. Galvin, Fast agglomeration of ultrafine
  hydrophobic particles using a high-internal-phase emulsion binder comprising
  permeable hydrophobic films, Industrial \& Engineering Chemistry Research 56
  (2017) 10658--10666.

\bibitem{2020SciPy}
P.~Virtanen, R.~Gommers, T.~E. Oliphant, M.~Haberland, T.~Reddy, D.~Cournapeau,
  E.~Burovski, P.~Peterson, W.~Weckesser, J.~Bright, S.~J. {van der Walt},
  M.~Brett, J.~Wilson, K.~J. Millman, N.~Mayorov, A.~R.~J. Nelson, E.~Jones,
  R.~Kern, E.~Larson, C.~J. Carey, {\.I}.~Polat, Y.~Feng, E.~W. Moore,
  J.~{VanderPlas}, D.~Laxalde, J.~Perktold, R.~Cimrman, I.~Henriksen, E.~A.
  Quintero, C.~R. Harris, A.~M. Archibald, A.~H. Ribeiro, F.~Pedregosa, P.~{van
  Mulbregt}, {SciPy 1.0 Contributors}, {{SciPy} 1.0: Fundamental algorithms for
  scientific computing in Python}, Nature Methods 17 (2020) 261--272.

\bibitem{Nelsen2006}
R.~Nelsen, An Introduction to Copulas, Springer, 2006.

\bibitem{Kruis1997}
F.~Kruis, K.~Kusters, The collision rate of particles in turbulent flow,
  Chemical Engineering Communications 158 (1997) 201--230.

\bibitem{Wang2014}
X.~Wang, X.~Feng, C.~Yang, Z.-S. Mao, Energy dissipation rates of {Newtonian}
  and {non-Newtonian} fluids in a stirred vessel, Chemical Engineering \&
  Technology 37 (2014) 1575--1582.

\bibitem{Tomasi1998}
C.~Tomasi, R.~Manduchi, Bilateral filtering for gray and color images, in:
  Proceedings of the sixth International Conference on Computer Vision (IEEE
  Cat. No. 98CH36271), IEEE, 1998, pp. 839--846.

\bibitem{Gonzales2010}
R.~C. Gonzales, R.~E. Woods, Digital Image Processing, Pearson Education, 2010.

\bibitem{Furat2019}
O.~Furat, T.~Lei{\ss}ner, K.~Bachmann, J.~Gutzmer, U.~A. Peuker, V.~Schmidt,
  Stochastic modeling of multidimensional particle properties using parametric
  copulas, Microscopy and Microanalysis 25 (2019) 720--734.

\bibitem{Al-Thyabat2006}
S.~Al-Thyabat, N.~Miles, An improved estimation of size distribution from
  particle profile measurements, Powder Technology 166 (2006) 152--160.

\bibitem{Held2014}
L.~Held, D.~S. Bov\'e, Applied Statistical Inference: Likelihood and Bayes,
  Springer, 2014.

\bibitem{barndorff1977}
O.~Barndorff-Nielsen, Exponentially decreasing distributions for the logarithm
  of particle size, Proceedings of the Royal Society of London. Series A 353
  (1977) 401--419.

\bibitem{Abramowitz1968}
M.~Abramowitz, I.~A. Stegun, Handbook of Mathematical Functions with Formulas,
  Graphs, and Mathematical Tables, Vol.~55, US Government printing office,
  1968.

\bibitem{Anderson2003}
T.~W. Anderson, An Introduction to Multivariate Statistical Analysis, J. Wiley
  \& Sons, 2003.

\bibitem{scott2015}
D.~W. Scott, Multivariate Density Estimation: Theory, Practice, and
  Visualization, J. Wiley \& Sons, 2015.

\bibitem{Robert2004}
C.~Robert, G.~Casella, Monte Carlo Statistical Methods, Springer, 2004.

\bibitem{Karp1980}
R.~M. Karp, An algorithm to solve the m $\times$ n assignment problem in
  expected time {O}(mn log n), Networks 10 (1980) 143--152.

\bibitem{Loaiza2019}
G.~Loaiza-Ganem, J.~P. Cunningham, The continuous {B}ernoulli: fixing a
  pervasive error in variational autoencoders, in: Proceedings of the 33rd
  International Conference on Neural Information Processing Systems, Curran
  Associates Inc., Red Hook, NY, USA, 2019, pp. 13287--13297.

\bibitem{Mudholkar1993}
G.~S. Mudholkar, D.~K. Srivastava, Exponentiated {W}eibull family for analyzing
  bathtub failure-rate data, IEEE Transactions on Reliability 42 (1993)
  299--302.

\bibitem{Tadikamalla1980}
P.~R. Tadikamalla, A look at the {B}urr and related distributions,
  International Statistical Review / Revue Internationale de Statistique 48
  (1980) 377--344.

\bibitem{Johnson1995}
N.~L. Johnson, S.~Kotz, N.~Balakrishnan, Continuous Univariate Distributions,
  Vol.~2, J. Wiley \& Sons, 1995.

\bibitem{Gupta2008}
R.~D. Gupta, R.~C. Gupta, Analyzing skewed data by power normal model, TEST 17
  (2008) 197--210.

\bibitem{Talukdar1991}
K.~K. Talukdar, W.~D. Lawing, Estimation of the parameters of the {R}ice
  distribution, The Journal of the Acoustical Society of America 89 (1991)
  1193--1197.

\bibitem{Siekierski1992}
K.~Siekierski, Comparison and evaluation of three methods of estimation of the
  {J}ohnson {SB} distribution, Biometrical Journal 34 (1992) 879--895.

\bibitem{Gunduz2020}
S.~G{\"u}nd{\"u}z, M.~{\c{C}}. Korkmaz, A new unit distribution based on the
  unbounded {J}ohnson distribution rule: the unit {J}ohnson {SU} distribution,
  Pakistan Journal of Statistics and Operations Research 16 (2020) 3.

\bibitem{furat2020stochastic}
O.~Furat, M.~Masuhr, F.~E. Kruis, V.~Schmidt, Stochastic modeling of
  classifying aerodynamic lenses for separation of airborne particles by
  material and size, Advanced Powder Technology 31~(6) (2020) 2215--2226.

\end{thebibliography}

\end{document}